\documentclass[acmsmall]{acmart}
\AtBeginDocument{%
  }

\setcopyright{cc}
\setcctype{by}
\acmJournal{PACMSE}
\acmYear{2026}
\acmVolume{3}
\acmNumber{ISSTA}
\acmArticle{ISSTA188}
\acmMonth{10}
\acmDOI{10.1145/3832279}

\acmSubmissionID{issta26main-p2238-p}
\received{2026-01-30}
\received[accepted]{2026-04-16}

\usepackage{amsmath}

\usepackage{amssymb}
\usepackage{amsfonts}
\usepackage{algorithmic}
\usepackage{graphicx}
\usepackage{textcomp}
\usepackage[ruled,vlined]{algorithm2e}

\usepackage{float}
\usepackage{booktabs}
\usepackage{multirow}
\usepackage{array} 
\usepackage{makecell}
\usepackage{threeparttable}
\usepackage{tcolorbox}
\usepackage{xcolor}
\usepackage[skip=3pt]{caption}

\begin{document}

\title{Contamination Means Overestimation? A Fine-Grained Empirical Study in Code Intelligence}

\author{Zhen Yang}
\affiliation{%
  \institution{Shandong University}
  \city{Qingdao}
  \country{China}
}
\email{zhenyang@sdu.edu.cn}

\author{Hongyi Lin}
\affiliation{%
  \institution{Shandong University}
  \city{Qingdao}
  \country{China}
}
\email{Hongyi.Lin@mail.sdu.edu.cn}

\author{Yifan He}
\affiliation{%
  \institution{Shandong University}
  \city{Qingdao}
  \country{China}
}
\email{yifan.he@mail.sdu.edu.cn}

\author{Junqi Wang}
\affiliation{%
  \institution{Shandong University}
  \city{Qingdao}
  \country{China}
}
\email{jqw\_sdu@163.com}

\author{Zeyu Sun}\authornote{Corresponding Author}
\affiliation{%
  \institution{National Key Laboratory of Space Integrated Information System, Institute of Software, Chinese Academy of Sciences}
  \city{Beijing}
  \country{China}
}
\email{zeyu.zys@gmail.com}

\author{Shuo Liu}
\affiliation{%
  \institution{City University of Hong Kong}
  \city{Hong Kong}
  \country{China}
}
\email{sliu273-c@my.cityu.edu.hk}

\author{Jie Xu}
\affiliation{%
  \institution{Shandong University}
  \city{Qingdao}
  \country{China}
}
\email{jiexu1@mail.sdu.edu.cn}

\author{Pengpeng Wang}
\affiliation{%
  \institution{Columbia University}
  \city{New York}
  \country{USA}
}
\email{pw2660@columbia.edu}

\author{Zhongxing Yu}\authornotemark[1]
\affiliation{%
  \institution{Shandong University}
  \city{Qingdao}
  \country{China}
}
\email{zhongxing.yu@sdu.edu.cn}
\author{Qingyuan Liang}
\affiliation{%
  \institution{Peking University}
  \city{Beijing}
  \country{China}
}
\email{liangqy@pku.edu.cn}

\renewcommand{\shortauthors}{Zhen Yang et al.}

\begin{abstract}
In recent years, code intelligence has gained increasing importance in the field of automated software engineering. Meanwhile, the widespread adoption of Pretrained Language Models (PLMs) and Large Language Models (LLMs) has raised concerns regarding data contamination and its potential impact on model performance evaluation. Previous studies mainly focused on sample-level contamination, ignoring partial contamination scenarios that are pervasive in code intelligence.
This paper fills this gap and presents a systematic empirical study to investigate the fine-grained data contamination on mainstream code tasks.
Our study involves diverse representative PLMs: RoBERTa and GPT-2, and LLMs: LLaMA and StarCoder, covering three major tasks: code translation, code generation, and code summarization, across two Programming Languages (PLs): Java and Python. We categorize contamination scenarios into four types according to the code intelligence practice, namely input-only, output-only, unpaired, and paired contamination settings, and construct corresponding experimental and control groups for exploration.

Experimental results show that, under the pre-training, fine-tuning, and inference paradigm adopted by PLMs, even deliberately injecting paired contamination does not lead to significant performance overestimation. But direct inference or small-scale fine-tuning uncovers the contamination effects. 
In contrast, LLMs with pre-training and inference paradigm are significantly affected by the paired contamination. Apart from the above, other contamination scenarios have no impact on both PLMs and LLMs.
Our findings challenge the conventional belief that contamination inevitably leads to performance overestimation, providing new insights into the evaluation and deployment of code intelligence models.
\end{abstract}




\maketitle

\section{Introduction}
\label{Introduction}
In recent years, code intelligence has been a critical research area in the software engineering field, significantly improving software development and maintenance automation~\cite{sun2025surveyneuralcodeintelligence,zhang2024unifyingperspectivesnlpsoftware,jiang2024survey,alon2018code2seq,alon2019code2vec,zhang2019novel}. This is mainly attributed to the advancement of Pre-trained Language Models (PLMs)~\cite{feng-etal-2020-codebert, guo2020graphcodebert,svyatkovskiy2020intellicode,lu2021codexglue,ahmad2021unified,wang2021codet5,wang2023codet5+,guo2022unixcoder} and Large Language Models (LLMs)~\cite{chen2021evaluating,touvron2023llama,nijkamp2022codegen,li2023starcoder,roziere2023code,lozhkov2024starcoder} on code, especially their effective learning on massive Natural Language (NL) and Programming Language (PL) corpora. 
Despite their success, researchers began to question whether PLMs or LLMs' high performances stem from their memorization of pre-training data, i.e., data contamination, 
and many related studies were proposed \cite{cao2024concerneddatacontaminationassessing,ramos2024large,riddell2024quantifying,magar2022data,golchin2024time}. Nevertheless, most of them concentrate on sample-level contamination, which means both input and output portions are leaked in pairs in PLMs/LLMs' pre-training corpora.  
However, contamination does not always show up in samples, especially in the code intelligence field. For example, when doing code translation from Java to ArkTS, a declarative PL for application development in the HarmonyOS ecosystem \cite{liu2024llm}. PLMs/LLMs may not be pre-trained on ArkTS, but have seen massive Java code, including the one to be translated. In this case, a contamination scenario concerning only input portions occurs. Other scenarios are also prevalent in code tasks, such as the unit test generation task. Perhaps both the focal methods and test cases have been pre-trained by PLMs/LLMs, but they are placed in different files on GitHub, leading to their unpaired learning during the pre-training stage of PLMs/LLMs. In this case, how does it affect PLMs/LLMs' performance on downstream tasks? As such, this work seeks to fill this important gap.

\textbf{Empirical Study:} To explore the contamination effects in a fine-grained manner, we conduct a comprehensive empirical study involving two PLMs and two LLMs on three representative code tasks, including code translation, code generation, and code summarization, encompassing two PLs, i.e., Java and Python.
In our analysis, we identify four typical types of data contamination scenarios commonly encountered in code intelligence: (1) input-only contamination, where only the input portion of the test sets appears in the pre-training corpora; (2) output-only contamination, where only the output portion is contaminated; (3) unpaired contamination, where both input and output components are seen in pre-training but not as matching pairs; and (4) paired contamination, where the exact input-output pairs from the test sets are found in the pre-training data of PLMs or LLMs.



As for experiments with PLMs, we adopt a \textbf{pre-training, fine-tuning, then inference paradigm} to simulate their practical usage. To circumvent uncontrolled data leakage, we perform pre-training from scratch~\cite{magar2022data,jiang2024does} and simulate contamination by injecting test data into the pre-training corpus according to different contamination settings. After fine-tuning on downstream tasks, we evaluate model performance on test sets and compare with the uncontaminated counterparts, thereby measuring the contamination effects. As for LLMs, we also follow their practical usage paradigm, i.e., \textbf{pre-training then inference}, to conduct the experiments. We use their publicly available pre-training corpora~\cite{touvron2023llama,li2023starcoder} to extract appropriate samples according to different contamination settings as contaminated testing sets. As for uncontaminated counterparts, we design a series of perturbation rules to modify the contaminated testing sets, making them unseen to LLMs~\cite{mastropaolo2023robustness} while keeping sample complexities the same. All experiments are repeated five times for statistical analysis, thereby making comparisons and discussing the effects of contamination.


\textbf{Main Findings:} (1) PLMs with pre-training, then fine-tuning and inference paradigm are surprisingly not significantly affected by any contamination settings, yielding only marginal fluctuations on CrystalBLEU/CodeBLEU for code translation and generation, and on BLEU/ROUGE-L for code summarization, over uncontaminated counterparts across three code tasks (\textit{p}-value$>$0.05). 
(2) Direct inference or small-scale fine-tuning uncovers the contamination effects on decoder-only PLMs, such as GPT-2. In particular, direct inference induces the most significant overestimation, reaching 15.57\%--65.62\% on CrystalBLEU and 8.75\%--314.80\% on CodeBLEU for code translation, 14.29\%--85.47\% on CrystalBLEU and 3.85\%--152.17\% on CodeBLEU for code generation, and 17.86\%--36.14\% on BLEU and 14.58\%--67.21\% on ROUGE-L for code summarization (\textit{p}-value$<$0.05). 
(3) Large-scale fine-tuning tends to eliminate the contamination effects of decoder-only PLMs by overwhelmingly learning with underlying knowledge and task alignment specifications.
(4) LLMs with pre-training then inference paradigm are only affected by paired contamination, achieving consistent overestimation of 5.62\%--21.82\% on CrystalBLEU/CodeBLEU for code generation, and 6.50\%--18.43\% on BLEU/ROUGE-L for code summarization (\textit{p}-value$<$0.05). (5) The above conclusions apply to both static- and dynamic-typed PLs, such as Java and Python.  
This paper makes the following contributions. All code and data can be found at~\cite{anonymous:online}.

(1) To the best of our knowledge, this is the first systematic study exploring fine-grained data contamination effects in code intelligence.

(2) We propose a series of reproducible methods to extract contaminated samples for three code tasks and four contamination settings from the pretraining corpora of open-source LLMs, facilitating future contamination-related research. 

(3) We conduct extensive experiments among diverse contamination settings, various PLMs/LLMs on multiple code tasks and programming languages. A series of findings are revealed in this work, shedding light on future research and practice.


    
    

    

\section{Experimental Setup}
We conduct the data contamination exploration on two mainstream models in code intelligence, i.e., Pre-trained Language Models (PLMs) and Large Language Models (LLMs). Towards each category, we experiment with multiple code-related tasks and varying contamination settings. 
In the following subsections, we first introduce our research questions, then present the corresponding explored models, tasks, and experimental methodologies.

\subsection{Research Questions}
\label{Research Questions}
This paper proposes the following four Research Questions (RQs).

\subsubsection{RQ1: Does Input-Only Contamination Lead to Performance Overestimation?}\label{RQ1: Does Input-Only Contamination Lead to Performance Overestimation?}
When constructing testing sets, only the input portion is included in the pre-training corpora of PLMs/LLMs, thereby simulating the input-only contamination. This is a prevalent scenario, and the Java-to-ArkTS translation is a typical example, as mentioned in Section \ref{Introduction}.

A prior study conducted by Jiang et al. \cite{jiang2024does} has claimed that the input-only and output-only contamination leads to performance overestimation for PLMs in Natural Language (NL) tasks, such as SQuAD \cite{rajpurkar2016squad} and MMLU \cite{hendrycks2020measuring}. However, NL tasks differ significantly from code intelligence tasks, and their experimental designs have several limitations. First, the NL tasks carried out in Jiang et al.'s study only involve English text, leading to a relatively high similarity between their inputs and outputs compared with code intelligence tasks that always contain both NLs and code from different PLs. Second, the experiments of Jiang et al. only make comparisons in terms of average performance among only three trials without conducting statistically significant tests, making it hard to eliminate the influence of randomness and determine whether the outperformance of data contamination is consistent and prominent. Third, Jiang et al. only conducted their experiments on one PLM, i.e., GPT-2, without considering different architectures of PLMs and other LLMs. 
Therefore, the input-only contamination is highly worthwhile to re-evaluate in the code intelligence domain.

\subsubsection{RQ2: Does Output-Only Contamination Lead to Performance Overestimation?}\label{RQ2: Does Output-Only Contamination Lead to Performance Overestimation?}
Contrary to the above setting, output-only contamination places the output portion of testing sets in the pre-training corpora of PLMs/LLMs. 
Similar to the previous setting, this scenario is also practical in code intelligence and differs significantly from that in NL tasks. Thus, resolving this RQ is also necessary.

\subsubsection{RQ3: Does Unpaired Contamination Lead to Performance Overestimation?}\label{RQ3: Does Unpaired Contamination Lead to Performance Overestimation?}
This is a kind of complete contamination of testing sets, but the input and output portions are not paired in the pre-training corpora of PLMs/LLMs. Recall the example we mentioned in Section \ref{Introduction}, when doing 
unit test case generation task, both focal methods and test cases 
may exist in the pre-training corpora of PLMs/LLMs, but their source files are split in the wild. Thus, these code samples will be pre-trained in PLMs/LLMs in a split manner. Considering the prevalence of this setting, unpaired input-output contamination is also meaningful and critical for exploration.
 
\subsubsection{RQ4: Does Paired Contamination Lead to Performance Overestimation?}\label{RQ4: Does Paired Contamination Lead to Performance Overestimation?}
As another kind of complete contamination, paired input-output contamination represents that both the input and output portions of testing sets have been seen by PLMs/LLMs, and they are concatenated in pairs during the pre-training stage. Many code intelligence tasks are very likely to involve this contamination, such as code generation and completion, as their inputs (i.e., NL or code prefix) are always closely attached in front of corresponding outputs (i.e., code snippets) in the wild, leading to their paired appearance in PLMs/LLMs' pre-training. Theoretically, paired contamination is the most likely type to cause performance overestimation, and the LLM part has been investigated in \cite{riddell2024quantifying,ramos2024large}, but PLMs with pre-training, fine-tuning, and inference paradigm have not been explored.

\subsection{Models under Test}
According to the wide acknowledgement within the community \cite{sun2025surveyneuralcodeintelligence,zhang2024unifyingperspectivesnlpsoftware,jiang2024survey}, the parameter sizes and training bases of Large Language Models (LLMs) are sufficiently large with emerging abilities; thus, they are frequently employed in practice using the pre-training-then-inference paradigm. In contrast, Pre-trained Language Models (PLMs) with much smaller scales are more commonly utilized in practice following the pre-train, fine-tune, then infer paradigm. Therefore, throughout this paper, both PLMs and LLMs are selected for experiments, thereby ensuring a comprehensive study and the alignment of experimental results and practice. 

\subsubsection{Pre-trained Language Models (PLMs):} Among this category, an encoder-only model, namely RoBERTa~\cite{liu2019roberta}, and a decoder-only model, namely GPT-2~\cite{radford2019language}, are selected for experiments, as they have covered the most representative architectures of PLMs. We select RoBERTa with 125M parameters and GPT-2 with 150M parameters for experiments according to the load of our computational resources.
To facilitate our research, we adopt one typical pre-training task for each PLM during its pre-training stage. RoBERTa is pre-trained with the Masked Language Modeling (MLM)~\cite{devlin2019bertpretrainingdeepbidirectional} task, while the Causal Language Modeling (CLM)~\cite{radford2018improving} task is used for GPT-2. Contamination injections are performed in the pre-training stage. Then, PLMs go through fine-tuning and inference stages to observe whether contamination induces performance overestimation compared with the uncontaminated counterparts.


\subsubsection{Large Language Models (LLMs):}
As for this category, we select one general LLM, namely LLaMA~\cite{touvron2023llama}, and one code LLM, namely StarCoder~\cite{li2023starcoder}. LLaMA is a family of multilingual LLMs with parameter sizes ranging from 7B to 65B. It is pre-trained on 1.4 trillion tokens of various open-source datasets, where 4.5\% of the training base is code extracted from GitHub \cite{GitHub70:online}. We chose version 33B without instruction fine-tuning for experiments. StarCoder is a Code LLM containing 15.5B parameters, trained on over 1 trillion code tokens. 
LLMs are experimented with in a pre-training then inference paradigm. Due to the massive scale and substantial training base of LLMs, retraining from scratch is impractical. We curate appropriate samples from their pre-training corpora for the contaminated testing set construction and experiments.
The reasons we consider the above two LLMs are threefold: First, they cover both general and code LLMs, making our exploration comprehensive. Second, they are two of the few open-source LLMs that also release their pre-training corpora, enabling us to construct contaminated testing sets and the uncontaminated counterparts from them. Third, they all went through a pre-training stage only, without instruction/preference tuning \cite{weifinetuned, ziegler2019fine}, making them suitable for pre-training, then the inference setting.

\subsection{Code Tasks and Metrics}
\label{Code Tasks and Metrics}
\subsubsection{Code Translation:} This task involves converting code from one PL to another while keeping the functionality the same, which is critical in scenarios such as migrating legacy systems and adopting modern frameworks. In this work, we focus on code translation from Java to C\#, and Python to Java for experiments, denoted as Java$\rightarrow$C\# and Python$\rightarrow$Java for simplicity. 

\subsubsection{Code Generation:} It is a process of automatically producing source code from a given NL requirement, aiming to reduce manual coding effort while ensuring correctness and efficiency. In this work, we adopt both Java and Python code generation for experiments, which is denoted as NL$\rightarrow$Java and NL$\rightarrow$Python for simplicity.

\subsubsection{Code Summarization:} This task focuses on generating an NL description for a given code snippet, which is crucial for improving code readability, maintenance, and collaboration among developers. Still, we carry out experiments on both Java and Python programs, which are denoted as Java$\rightarrow$NL and Python$\rightarrow$NL for simplicity.

\subsubsection{Metrics:}
To automatically assess the quality of the outputs for the above code tasks, we adopt task-specific evaluation metrics. For \textbf{code translation} and \textbf{code generation}, whose outputs are source code, we adopt CrystalBLEU~\cite{eghbali2022crystalbleu} and CodeBLEU~\cite{ren2020codebleu}. CrystalBLEU mitigates the bias of vanilla BLEU by ignoring the most frequent shared $n$-grams in code corpora, providing a more discriminative measure of code similarity. CodeBLEU augments $n$-gram overlap with weighted token matching, AST matching, and data-flow matching, capturing both lexical and syntactic/semantic correctness of the generated code.
For \textbf{code summarization}, whose outputs are natural language descriptions, we adopt BLEU~\cite{papineni2002bleu} and ROUGE-L~\cite{lin2004rouge}. BLEU measures the precision of $n$-gram overlap between the generated summary and the reference, while ROUGE-L measures the recall based on the longest common subsequence, capturing fluency and content coverage of the generated summaries, respectively.
By combining these complementary metrics for each task, we provide a more balanced and holistic assessment of generation quality from both precision and recall perspectives.

\subsection{Datasets}
\label{Datasets}

\subsubsection{Dataset preparation for PLMs:}
Since all included code tasks are divided into a Java side and a Python side as mentioned in Section \ref{Code Tasks and Metrics}, we use the Java and Python portions of the widely adopted CodeSearchNet dataset~\cite{husain2019codesearchnet} to pre-train PLMs, respectively, for follow-up exploration accordingly. 
The Java portion contains over 160,000 Java function snippets, while the Python portion includes approximately 250,000 Python function snippets. 

Afterwards, to smoothly carry out experiments on each code task, we carefully select a series of typical datasets.
For the \textbf{code translation} task, we adopt the CodeTrans dataset~\cite{lu2021codexglue} for experiments on the Java-to-C\# translation, which includes 10,300 training samples, 500 validation samples, and 1,000 test samples. As for the Python-to-Java translation, we use the AVATAR dataset \cite{ahmad-etal-2023-avatar}, containing 60,138 training samples, 476 validation samples, and 1,906 testing samples.
For the \textbf{code generation} task, we use the CONCODE dataset~\cite{iyer2018mapping} for NL-to-Java fine-tuning and inference, which contains 100,000 training samples and 2,000 validation and testing samples. For NL-to-Python code generation, we conduct experiments on the Gretel AI dataset \cite{gretel-text-to-python-fintech-v1}, which consists of 22,500 training samples, 2,500 validation samples, and 2,500 test samples. 
For the \textbf{code summarization} task, TL-CodeSum dataset \cite{ijcai2018p314} is selected for experiments on summarizing Java code via NL, consisting of 55,767 training samples and 6,970 validation and testing samples, respectively. In addition, the code-docstring-corpus dataset \cite{miceli-barone-sennrich-2017-parallel} is chosen for Python-targeted code summarization, where 109,108 samples are for training, and 2,000 samples are for validation and testing, respectively.
Since the above datasets are open-source and the corresponding illustrations are detailed in their original papers, we do not list their specific statistics here to stay within page limits.
To restrict the experimental scales while maintaining the validity of the whole research, we limit the size of the training set to no more than 100,000 samples, the validation set to no more than 500 samples, and the testing set to 1,000 samples at most; when their original volumes exceed the above thresholds, we make corresponding random sampling. Detailed experimental settings and methodologies are presented in Section \ref{Research Methodology}.  



\subsubsection{Dataset preparation for LLMs:} To construct contaminated datasets and their uncontaminated counterparts for LLMs, we curate their released pre-training corpora, i.e., Google BigQuery for LLaMA and The Stack (v1.2) for StarCoder. The GitHub dataset available on Google BigQuery constitutes 4.5\% of the pre-training corpus of LLaMA, accounting for 328 GB. As the main training base of StarCoder, Stack (v1.2) consists of 6.4 TB of source code in 384 PLs. We construct each code task's experimental data from these two corpora, and the details can be found in Section \ref{Research Methodology}.
The dataset statistics are reported per RQ in Tables~\ref{tab:data-rq1}--\ref{tab:data-rq4}, where \textbf{Avg}, \textbf{Std}, and \textbf{Med} denote the average, standard deviation, and median of the token counts per sample, respectively; \textbf{Lang} indicates the language to which each input/output portion belongs; and \textbf{Perturbance count (ratio)} represents the average number and proportion of perturbed tokens. Throughout these tables, Ja, Py, and NL stand for Java, Python, and Natural Language, and underlines mark the contaminated portion of each RQ's experimental group.

\subsection{Research Methodology}
\label{Research Methodology}
This section introduces specific experimental procedures that investigate the RQs proposed in Section \ref{Research Questions}.
In principle, both PLMs and LLMs should be evaluated on all three code tasks, i.e., code translation, generation, and summarization. PLMs indeed are the case, as their scales are feasible for pre-training from scratch, making it easier to precisely control the contamination settings. Nonetheless, since LLMs are too huge to be retrained from scratch, we cannot modify their pre-training corpora but have to proactively construct contaminated test sets according to their pre-training corpora. In this case, experimental settings for certain code tasks may not be satisfied and dropped.
To make the whole narration clearer, we separately elaborate on the research methodology for PLMs and LLMs in each RQ. If certain code tasks are dropped in LLMs' experiments, we will provide detailed explanations accordingly. To make it clearer to readers about the dataset preparation for both the control and experimental groups, we list the associated illustrations in \cite{anonymous:online}. 

\subsubsection{RQ1: Does Input-Only Contamination Lead to Performance Overestimation?}\label{RQ1: method} \hfill \\
\textbf{(1) PLMs (Pre-train, fine-tune, then infer):}
For the Java side, \textbf{code translation} task (Java$\rightarrow$C\#) of PLMs dictating each Java code as input while using each corresponding C\# code as output. When constructing the uncontaminated control group, we directly use our prepared 160,000 Java samples for PLMs to pre-train from scratch, then fine-tune and infer on the CodeTrans dataset. Towards the contaminated experimental group, for each sample in CodeTrans's testing set, we inject its input portion (i.e., Java) into the PLMs' pre-training corpus while leaving its output portion (i.e., C\#) intact. To eliminate the bias of pre-training size brought about by the contamination injection, an equal number of randomly selected samples is removed from the pre-training corpus in advance in the experimental group. Afterwards, PLMs are pre-trained on the contaminated corpora, then fine-tune and infer on CodeTrans for performance comparison with the corresponding control group.
As for the \textbf{code generation} task (NL$\rightarrow$Java), the inputs are NLs, while the outputs are Java code snippets, while the \textbf{code summarization} task (Java$\rightarrow$NL) requires the Java code as inputs and NLs as outputs. We follow the same procedure as the code translation task to carry out the experiments, except that the fine-tuning and inference stages are supported by CONCODE and TL-CodeSum datasets, respectively. 

In addition, for the Python-to-Java code translation, NL-to-Python code generation, and Python-to-NL code summarization tasks, we use the 250,000 Python samples for PLMs' pre-training, while harnessing AVATAR, Gretel AI's dataset, and code-docstring-corpus for their respective fine-tuning and inference. All other procedures are the same as the above Java side. 

\textbf{(2) LLMs (Pre-train, then infer):}
Starting from the Java side and its \textbf{code translation} task (Java$\rightarrow$C\#), we first manually select 100 Java functions from both Google BigQuery and The Stack datasets as input portions of the contaminated experimental group.
Afterwards, we manually migrate the above Java functions into C\#, forming code translation pairs, thereby ensuring the output portions are unseen to LLMs. To guarantee the migration quality, we follow Xue et al. \cite{xue2025classeval}'s procedure and adapt their methodology into our manual migration, such as line-wise translation and keeping the naming conventions of C\#. In practice, two authors with more than three years of Java, Python, and C\# development experience are involved in this procedure above. One is responsible for implementation, while the other is for double-checking. As such, the experimental group of Java-to-C\# code translation can be constructed.   
As for the uncontaminated control group, we perturb the above contaminated samples manually to weaken the model's ability to memorize samples from pretraining directly.
Besides, perturbations should be consistently applied to both samples' inputs and outputs, making sure they are still paired after perturbation. Meanwhile, to guarantee that the task complexities (e.g., cyclomatic complexity~\cite{mccabe1976complexity} and code length) are unchanged after perturbation for fair comparisons, only sub-token-level replacement in user-defined identifiers, e.g., function/variable names, is permitted (e.g., perturbing \texttt{fieldName} to \texttt{fieldIdentifier}), while leaving control flow structures, data types, and core logic intact.
Following this protocol, we define the \textbf{perturbation rate} of a sample as the proportion of user-defined identifiers among all its tokens. Because the number of such identifiers naturally varies across code samples and across tasks, the perturbation rate also varies, which explains the differing rates reported in Tables~\ref{tab:data-rq1}--\ref{tab:data-rq4}.

The two authors participate in this procedure again, one is responsible for the perturbation, while the other double-checks the perturbation results to ensure the perturbation satisfies the consistency requirement mentioned above.
As such, we construct both the experimental and control groups for the Java-to-C\# code translation task. Finally, we use LLMs to directly conduct inference on their associated experimental and control groups for performance difference comparison.

Regarding the \textbf{code summarization} task (Java$\rightarrow$NL), we still adopt the above-extracted 100 Java functions as contaminated input portions, and manually craft their corresponding NLs, thereby making the output portions (i.e., NLs) unseen to LLMs and forming the code summarization pairs of the contaminated experimental group. Similarly, we adhere to a series of standards when crafting NLs. 1) NLs must reflect the primary purpose of code snippets. 2) Limit NLs to single sentences. 3) Focus on the high-level behaviors rather than low-level implementation details. 
The uncontaminated control group follows the same perturbation procedure as above for construction, thereby comparing performance with that of the experimental group. 
As for code tasks concerning Python, such as Python-to-Java code translation and Python-to-NL code summarization, we follow the same procedure to construct and compare both the experimental and control groups as above. 

\textbf{Code generation task} (NL$\rightarrow$Java/Python) for LLMs' experiments is ignored in this RQ, because it requires finding standalone NLs describing code functionalities without associated codes as contamination sources, in LLMs' pre-training corpora, which are extremely scarce. To make readers learn more about the constructed datasets in both groups, we present Table~\ref{tab:data-rq1} to illustrate their statistics.

\begin{table}[h]
\scriptsize
\setlength{\abovecaptionskip}{0cm}
\caption{Testing Sets Constructed for LLMs' Experiments in RQ1 (Input-Only Contamination)}
\label{tab:data-rq1}
\begin{tabular}{llccccccccc}
\toprule
\multirow{2}{*}{\textbf{LLM}} & \multicolumn{2}{c}{\multirow{2}{*}{\textbf{Task (In / Out)}}} & \multirow{2}{*}{\textbf{Lang}} & \multicolumn{3}{c}{\textbf{Experimental Group}} & \multicolumn{3}{c}{\textbf{Control Group}} & \multirow{2}{*}{\begin{tabular}[c]{@{}c@{}}Perturbance\\Count (Ratio)\end{tabular}} \\ \cline{5-10}
 &  &  &  & \textbf{Avg} & \textbf{Std} & \textbf{Med} & \textbf{Avg} & \textbf{Std} & \textbf{Med} & \\ \midrule
\multirow{8}{*}{LLaMA} & \multicolumn{2}{c}{\multirow{2}{*}{Code Translation (\underline{Ja}$\rightarrow$C\#)}} & Ja & 62.47 & 27.09 & 55.50 & 62.84 & 26.88 & 57.00 & 9.73 (15.33\%) \\
 & \multicolumn{2}{c}{} & C\# & 58.86 & 28.18 & 51.00 & 59.63 & 28.23 & 51.50 & 9.27 (15.49\%) \\ \cline{4-11}
 & \multicolumn{2}{c}{\multirow{2}{*}{Code Translation (\underline{Py}$\rightarrow$Ja)}} & Py & 80.61 & 32.43 & 73.50 & 81.64 & 31.96 & 75.50 & 11.53 (13.80\%) \\
 & \multicolumn{2}{c}{} & Ja & 123.10 & 58.71 & 115.50 & 123.09 & 59.48 & 108.00 & 17.96 (14.95\%) \\ \cline{4-11}
 & \multicolumn{2}{c}{\multirow{2}{*}{Code Summarization (\underline{Ja}$\rightarrow$NL)}} & Ja & 62.47 & 27.09 & 55.50 & 62.84 & 26.88 & 57.00 & 9.73 (15.33\%) \\
 & \multicolumn{2}{c}{} & NL & 30.99 & 7.56 & 30.00 & 30.36 & 7.31 & 30.00 & 15.54 (48.58\%) \\ \cline{4-11}
 & \multicolumn{2}{c}{\multirow{2}{*}{Code Summarization (\underline{Py}$\rightarrow$NL)}} & Py & 80.61 & 32.43 & 73.50 & 81.64 & 31.96 & 75.50 & 11.53 (13.80\%) \\
 & \multicolumn{2}{c}{} & NL & 27.10 & 6.33 & 27.00 & 26.45 & 5.49 & 26.50 & 9.39 (33.85\%) \\ \hline
\multirow{8}{*}{StarCoder} & \multicolumn{2}{c}{\multirow{2}{*}{Code Translation (\underline{Ja}$\rightarrow$C\#)}} & Ja & 52.30 & 21.20 & 48.00 & 52.82 & 21.01 & 49.00 & 8.21 (15.70\%) \\
 & \multicolumn{2}{c}{} & C\# & 58.86 & 28.18 & 51.00 & 59.63 & 28.23 & 51.50 & 9.27 (15.49\%) \\ \cline{4-11}
 & \multicolumn{2}{c}{\multirow{2}{*}{Code Translation (\underline{Py}$\rightarrow$Ja)}} & Py & 66.06 & 26.43 & 63.00 & 66.37 & 26.96 & 62.50 & 11.41 (17.86\%) \\
 & \multicolumn{2}{c}{} & Ja & 108.04 & 66.28 & 92.50 & 108.08 & 67.90 & 90.00 & 19.42 (17.69\%) \\ \cline{4-11}
 & \multicolumn{2}{c}{\multirow{2}{*}{Code Summarization (\underline{Ja}$\rightarrow$NL)}} & Ja & 52.30 & 21.20 & 48.00 & 52.82 & 21.01 & 49.00 & 8.21 (15.70\%) \\
 & \multicolumn{2}{c}{} & NL & 28.02 & 6.18 & 27.50 & 28.26 & 6.23 & 28.00 & 14.26 (50.89\%) \\ \cline{4-11}
 & \multicolumn{2}{c}{\multirow{2}{*}{Code Summarization (\underline{Py}$\rightarrow$NL)}} & Py & 66.06 & 26.43 & 63.00 & 66.44 & 26.79 & 62.50 & 12.22 (19.26\%) \\
 & \multicolumn{2}{c}{} & NL & 25.23 & 4.71 & 25.00 & 24.51 & 4.37 & 24.00 & 10.24 (40.54\%) \\ \bottomrule
\end{tabular}
\end{table}

\subsubsection{RQ2: Does Output-Only Contamination Lead to Performance Overestimation?}\label{RQ2: method} \hfill \\
\textbf{(1) PLMs (Pre-train, fine-tune, then infer):}
For the Java side, we follow the same experimental procedures as in RQ1, except that the contamination is injected on the output side rather than the input side. Specifically, for the Java-to-C\# \textbf{code translation} task assessed on CodeTrans, we inject each of their testing samples' output portion (i.e., C\#) into the pre-training corpora of PLMs, while keeping the corresponding input portion (i.e., Java) intact. Similarly, for the NL-to-Java \textbf{code generation} and Java-to-NL \textbf{code summarization} tasks, we contaminate the pre-training corpus with the Java code and NL from their corresponding test sets, respectively, as they serve as the output side. The same strategy is also applied to Python-side evaluations, such as the Python-to-Java code translation, NL-to-Python code generation, and Python-to-NL code summarization tasks. 

\textbf{(2) LLMs (Pre-train, then infer):}
Still beginning with the Java side, the \textbf{code translation} task (Java$\rightarrow$C\#) follows a similar procedure to RQ1 for both experimental and control group construction and experimentation. The only difference is that we first randomly collect 100 C\# code snippets from Google BigQuery and The Stack, then manually translate them into Java. As such, the input portion (i.e., Java) of this test set is unseen to LLMs, while the output portion (i.e., C\#) is contaminated.
For the \textbf{code generation} task (NL$\rightarrow$Java), we first randomly collect 100 Java code snippets from Google BigQuery and The Stack, then manually translate their functionalities into NL descriptions, making NL-to-Java pairs with the unseen inputs and contaminated outputs to LLMs. Other procedures, such as experimental and control groups construction, as well as experimentation, are the same as those in RQ1.
The same procedure above is applied to code tasks on the Python side (Python$\rightarrow$Java and NL$\rightarrow$Python).
Besides, we do not conduct the \textbf{code summarization} task (Java/Python$\rightarrow$NL) for LLMs in RQ2 for the same reason as RQ1's code generation task. 
The detailed statistics of the constructed datasets for RQ2 are presented in Table~\ref{tab:data-rq2}.

\begin{table}[h]
\scriptsize
\setlength{\abovecaptionskip}{0cm}
\caption{Testing Sets Constructed for LLMs' Experiments in RQ2 (Output-Only Contamination)}
\label{tab:data-rq2}
\begin{tabular}{llccccccccc}
\toprule
\multirow{2}{*}{\textbf{LLM}} & \multicolumn{2}{c}{\multirow{2}{*}{\textbf{Task (In / Out)}}} & \multirow{2}{*}{\textbf{Lang}} & \multicolumn{3}{c}{\textbf{Experimental Group}} & \multicolumn{3}{c}{\textbf{Control Group}} & \multirow{2}{*}{\begin{tabular}[c]{@{}c@{}}Perturbance\\Count (Ratio)\end{tabular}} \\ \cline{5-10}
 &  &  &  & \textbf{Avg} & \textbf{Std} & \textbf{Med} & \textbf{Avg} & \textbf{Std} & \textbf{Med} & \\ \midrule
\multirow{8}{*}{LLaMA} & \multicolumn{2}{c}{\multirow{2}{*}{Code Translation (Ja$\rightarrow$\underline{C\#})}} & Ja & 93.83 & 45.31 & 84.00 & 90.26 & 42.93 & 81.00 & 20.27 (20.21\%) \\
 & \multicolumn{2}{c}{} & C\# & 72.89 & 29.81 & 80.00 & 71.30 & 28.93 & 75.50 & 15.14 (19.50\%) \\ \cline{4-11}
 & \multicolumn{2}{c}{\multirow{2}{*}{Code Translation (Py$\rightarrow$\underline{Ja})}} & Py & 77.21 & 42.01 & 68.50 & 75.36 & 38.99 & 66.00 & 17.58 (21.23\%) \\
 & \multicolumn{2}{c}{} & Ja & 77.14 & 32.83 & 72.50 & 76.91 & 32.60 & 73.00 & 12.75 (17.50\%) \\ \cline{4-11}
 & \multicolumn{2}{c}{\multirow{2}{*}{Code Generation (NL$\rightarrow$\underline{Ja})}} & NL & 90.49 & 22.59 & 86.00 & 99.22 & 29.07 & 95.00 & 27.58 (30.15\%) \\
 & \multicolumn{2}{c}{} & Ja & 62.47 & 27.09 & 55.50 & 62.84 & 26.88 & 57.00 & 9.73 (15.33\%) \\ \cline{4-11}
 & \multicolumn{2}{c}{\multirow{2}{*}{Code Generation (NL$\rightarrow$\underline{Py})}} & NL & 206.36 & 38.73 & 203.00 & 206.10 & 40.17 & 198.00 & 30.72 (14.89\%) \\
 & \multicolumn{2}{c}{} & Py & 80.61 & 32.43 & 73.50 & 82.00 & 31.99 & 76.00 & 11.54 (14.49\%) \\ \hline
\multirow{8}{*}{StarCoder} & \multicolumn{2}{c}{\multirow{2}{*}{Code Translation (Ja$\rightarrow$\underline{C\#})}} & Ja & 51.38 & 34.94 & 38.50 & 51.47 & 34.86 & 38.50 & 16.36 (31.84\%) \\
 & \multicolumn{2}{c}{} & C\# & 32.92 & 18.58 & 29.50 & 32.91 & 20.61 & 30.50 & 10.52 (35.59\%) \\ \cline{4-11}
 & \multicolumn{2}{c}{\multirow{2}{*}{Code Translation (Py$\rightarrow$\underline{Ja})}} & Py & 64.03 & 43.40 & 53.50 & 64.32 & 45.27 & 55.00 & 15.43 (24.00\%) \\
 & \multicolumn{2}{c}{} & Ja & 54.92 & 33.16 & 47.50 & 54.55 & 34.60 & 47.50 & 12.95 (23.58\%) \\ \cline{4-11}
 & \multicolumn{2}{c}{\multirow{2}{*}{Code Generation (NL$\rightarrow$\underline{Ja})}} & NL & 86.77 & 23.10 & 81.00 & 96.66 & 25.78 & 90.50 & 26.66 (30.72\%) \\
 & \multicolumn{2}{c}{} & Ja & 52.30 & 21.20 & 48.00 & 52.82 & 21.01 & 49.00 & 8.21 (15.70\%) \\ \cline{4-11}
 & \multicolumn{2}{c}{\multirow{2}{*}{Code Generation (NL$\rightarrow$\underline{Py})}} & NL & 216.50 & 44.25 & 206.00 & 222.01 & 45.45 & 209.00 & 22.67 (10.47\%) \\
 & \multicolumn{2}{c}{} & Py & 66.06 & 26.43 & 63.00 & 67.37 & 30.24 & 65.00 & 20.08 (30.40\%) \\ \bottomrule
\end{tabular}
\end{table}

\subsubsection{RQ3: Does Unpaired Contamination Lead to Performance Overestimation?}\label{RQ3:method} \hfill \\
\textbf{(1) PLMs (Pre-train, fine-tune, then infer):}
To conduct experiments with PLMs under this setting, we follow all the procedures described in RQ1. The key difference is: when constructing contaminated experimental groups, we include not only the input portions of downstream tasks but also the corresponding output portions for contaminating PLMs' pre-training corpora. For example, in the \textbf{code translation} task (Java$\rightarrow$C\#), we separate the Java and C\# portions of each sample in the CodeTrans test set and randomly inject them into the pre-training corpora of PLMs, thereby constructing the contaminated experimental group. 
The same procedure is applied to \textbf{code generation} (NL$\rightarrow$Java) and \textbf{summarization} (Java$\rightarrow$NL) tasks with their respective datasets on the Java side as well as to the equivalent Python-side experiments.

\textbf{(2) LLMs (Pre-train, then infer):} 
Constructing contaminated experimental groups in this RQ is relatively difficult compared with previous ones, as it requires both input and output portions of testing sets explicitly existing in the pre-training corpora of LLMs, but as independent samples. For \textbf{code generation} and \textbf{summarization} tasks, although finding independent code snippets in LLMs' pre-training corpora is easy, it is challenging to find an independent NL that coincidentally describes a code functionality but is not attached to a code snippet, because code and comments are normally closely attached in the wild. Therefore, we ignore the above two tasks for LLMs in the unpaired contamination setting and focus on the \textbf{code translation} task only.

Considering that manually finding those PL pairs (e.g., Java-C\# and Python-Java) in the tremendous pre-training corpora is extremely time-consuming, we propose a series of filtering rules to substantially shrink the searching range. 1) Functions of different PLs with identical function names normally carry the same functionalities. Thus, we first use \texttt{tree-sitter} to parse their function names, then remove special symbols (e.g., \texttt{\_}) to keep letters only, and switch them all to lowercase for the initial cross-PL function pair construction. 2) For each pair of cross-PL functions, we count some key features from their Abstract Syntax Trees (ASTs) for complexity comparison, thereby proceeding with further filtering. To be specific, we count their total number of AST nodes, statement nodes, expression nodes, and function-call nodes as key features. Function pairs on the ratios of all kinds of node counts within the range of $[0.5, 2.0]$ are kept; otherwise, they are discarded. 3) For the remaining function pairs, we extract their keywords (e.g., \texttt{while}, \texttt{if}, and \texttt{+}) and compute the Generalized Jaccard Index (GJI) \cite{zhang2019multi} as below for semantic-level filtering:
\begin{gather} 
    \label{eq:sim_feat} 
    \mathcal{S}_{\text{feat}}(x_s, x_t) 
    = \frac{\sum_{k \in \mathcal{V}} \min(C_s[k], C_t[k])}{\sum_{k \in \mathcal{V}} \max(C_s[k], C_t[k])} 
    \end{gather} 
where $x_s$ and $x_t$ denote the source and target functions, $\mathcal{V}$ represents the monitored keyword set, and $C_{\{s,t\}}[k]$ is the count of keyword $k$ in the function $x_s$ or $x_t$. Function pairs with $\mathcal{S}_{\text{feat}}>0.6$ are kept. 4) Finally, we further conduct an abstractive execution flow comparison based on ASTs. In detail, we extract the AST nodes in 2) with pre-order traversal to construct node sequences and compute the Longest Common Subsequence (LCS) \cite{bergroth2000survey} between function pairs:
\begin{gather} 
    \label{eq:sim_seq} 
    \mathcal{S}_{\text{seq}}(x_s, x_t) 
    = \frac{\big|\textsc{LCS}(Q_s, Q_t)\big|}{\max(|Q_s|, |Q_t|)} 
    \end{gather} 
where $Q_s$ and $Q_t$ are the node sequences of the source and target functions, and $|\cdot|$ denotes the sequence length. Function pairs with $\mathcal{S}_{\text{seq}}>0.6$ are kept.

Initially, we obtain over 60M Python-Java and 90M Java-C\# function pairs in Google BigQuery and 29M and 40M corresponding function pairs in Stack after the above step 1). After the following automated filtering steps 2)-4), we retain a much smaller and higher-quality candidate pool with 50,952 Python-Java and 60,900 Java-C\# function pairs from Stack, while 29,318 and 13,462 corresponding function pairs from Google BigQuery
for follow-up manual screening. As can be seen, the manual effort is substantially reduced.
Ultimately, for each code translation task, we manually select 100 function pairs to construct the contaminated experimental group. As for the uncontaminated control group construction and subsequent experiments, we follow the same procedure as previous RQs. 
Detailed statistics are shown in Table~\ref{tab:data-rq3}.

\begin{table}[h]
\scriptsize
\setlength{\abovecaptionskip}{0cm}
\caption{Testing Sets Constructed for LLMs' Experiments in RQ3 (Unpaired Contamination)}
\label{tab:data-rq3}
\begin{tabular}{llccccccccc}
\toprule
\multirow{2}{*}{\textbf{LLM}} & \multicolumn{2}{c}{\multirow{2}{*}{\textbf{Task (In / Out)}}} & \multirow{2}{*}{\textbf{Lang}} & \multicolumn{3}{c}{\textbf{Experimental Group}} & \multicolumn{3}{c}{\textbf{Control Group}} & \multirow{2}{*}{\begin{tabular}[c]{@{}c@{}}Perturbance\\Count (Ratio)\end{tabular}} \\ \cline{5-10}
 &  &  &  & \textbf{Avg} & \textbf{Std} & \textbf{Med} & \textbf{Avg} & \textbf{Std} & \textbf{Med} & \\ \midrule
\multirow{4}{*}{LLaMA} & \multicolumn{2}{c}{\multirow{2}{*}{Code Translation (\underline{Ja}$\rightarrow$\underline{C\#})}} & Ja & 60.09 & 61.00 & 38.50 & 59.94 & 60.95 & 37.50 & 12.41 (20.31\%) \\
 & \multicolumn{2}{c}{} & C\# & 61.13 & 68.37 & 40.50 & 61.13 & 68.50 & 39.00 & 13.20 (21.25\%) \\ \cline{4-11}
 & \multicolumn{2}{c}{\multirow{2}{*}{Code Translation (\underline{Py}$\rightarrow$\underline{Ja})}} & Py & 29.34 & 16.59 & 25.00 & 29.23 & 15.74 & 25.00 & 5.50 (21.01\%) \\
 & \multicolumn{2}{c}{} & Ja & 42.78 & 20.81 & 38.00 & 42.45 & 20.08 & 38.50 & 9.33 (20.62\%) \\ \hline
\multirow{4}{*}{StarCoder} & \multicolumn{2}{c}{\multirow{2}{*}{Code Translation (\underline{Ja}$\rightarrow$\underline{C\#})}} & Ja & 38.62 & 28.29 & 27.50 & 38.97 & 28.57 & 27.00 & 5.48 (14.19\%) \\
 & \multicolumn{2}{c}{} & C\# & 38.84 & 28.86 & 28.00 & 39.15 & 28.96 & 28.50 & 5.82 (14.98\%) \\ \cline{4-11}
 & \multicolumn{2}{c}{\multirow{2}{*}{Code Translation (\underline{Py}$\rightarrow$\underline{Ja})}} & Py & 30.09 & 13.26 & 27.00 & 30.63 & 14.60 & 31.00 & 8.08 (29.05\%) \\
 & \multicolumn{2}{c}{} & Ja & 38.51 & 17.39 & 34.00 & 39.54 & 18.11 & 35.50 & 8.33 (21.63\%) \\ \bottomrule
\end{tabular}
\end{table}

\subsubsection{RQ4: Does Paired Contamination Lead to Performance Overestimation?}\label{RQ4:method} \hfill \\
\textbf{(1) PLMs (Pre-train, fine-tune, then infer):}
Regarding the Java side, \textbf{code translation} task (Java$\rightarrow$C\#) under this setting continues following all the procedures described in RQ1. The key difference lies in the construction of the contaminated experimental group. We concatenate the input (i.e., Java) and output (i.e., C\#) portions of each sample in the CodeTrans test set using a separator token \texttt{<SEP>}, and inject these concatenated samples into the pre-training corpora of PLMs. We adopt \texttt{<SEP>} to distinguish different PLs, due to its wide usage in practice for bimodal pre-training~\cite{feng-etal-2020-codebert,wang2021codet5,zhu2024grammart5}. 
The same procedures are applied to the \textbf{code generation} and \textbf{summarization} tasks of the Java side as well as to the corresponding Python-related experiments.

\textbf{(2) LLMs (Pre-train, then infer):}
On the Java side, to construct the contaminated experimental group for the \textbf{code generation} task (NL$\rightarrow$Java) under this setting, we examine each Java sample in BigQuery and Stack using \texttt{tree-sitter} to determine whether the NL description appears immediately before the function signature. After that, we perform manual filtering to select 100 samples whose NL descriptions are accurate and informative enough to be used for code generation.
For the \textbf{code summarization} task (Java$\rightarrow$NL) under this setting, a paired contaminated sample should be an ordered concatenation of NL and Java code. However, it is rare to find a comment that appears after the function body both in the wild and LLMs' pre-training corpora. Therefore, we instead extract comments following the function signatures as desired NLs via \texttt{tree-sitter}, along with these functions, to construct the paired contamination samples. The reasons are two-fold: 1) Such comments often encapsulate the summarization of the function. 2) Such comments lie in front of their corresponding function body, largely satisfying our predefined settings. We then manually inspect the samples to determine whether the comments can serve as valid summaries, curating 100 proper samples for the experimental group.
The uncontaminated control group construction of the above two code tasks and their follow-up experiments are carried out following the same procedures described in RQ1. The aforementioned process is consistently applied to the Python-related tasks. 

In practice, code snippets with equivalent functionalities but written in different PLs are rarely placed in the same source files. As a result, it is difficult to find cross-PL program pairs with equivalent logic that appear as a sample in the pre-training corpora of LLMs. For this reason, we ignore the \textbf{code translation} task for both the Java and Python sides in the experiments with LLMs.
The detailed statistics of the constructed datasets for RQ4 are presented in Table~\ref{tab:data-rq4}.

\begin{table}[h]
\scriptsize
\setlength{\abovecaptionskip}{0cm}
\caption{Testing Sets Constructed for LLMs' Experiments in RQ4 (Paired Contamination)}
\label{tab:data-rq4}
\begin{tabular}{llccccccccc}
\toprule
\multirow{2}{*}{\textbf{LLM}} & \multicolumn{2}{c}{\multirow{2}{*}{\textbf{Task (In / Out)}}} & \multirow{2}{*}{\textbf{Lang}} & \multicolumn{3}{c}{\textbf{Experimental Group}} & \multicolumn{3}{c}{\textbf{Control Group}} & \multirow{2}{*}{\begin{tabular}[c]{@{}c@{}}Perturbance\\Count (Ratio)\end{tabular}} \\ \cline{5-10}
 &  &  &  & \textbf{Avg} & \textbf{Std} & \textbf{Med} & \textbf{Avg} & \textbf{Std} & \textbf{Med} & \\ \midrule
\multirow{8}{*}{LLaMA} & \multicolumn{2}{c}{\multirow{2}{*}{Code Generation (\underline{NL$\rightarrow$Ja})}} & NL & 47.75 & 46.03 & 32.00 & 47.99 & 45.27 & 31.50 & 9.20 (18.87\%) \\
 & \multicolumn{2}{c}{} & Ja & 63.70 & 46.04 & 47.50 & 63.94 & 46.47 & 48.00 & 12.32 (19.04\%) \\ \cline{4-11}
 & \multicolumn{2}{c}{\multirow{2}{*}{Code Generation (\underline{NL$\rightarrow$Py})}} & NL & 50.94 & 41.53 & 37.00 & 52.62 & 43.63 & 37.00 & 6.42 (14.94\%) \\
 & \multicolumn{2}{c}{} & Py & 73.08 & 52.15 & 60.50 & 74.35 & 56.33 & 62.00 & 18.46 (25.26\%) \\ \cline{4-11}
 & \multicolumn{2}{c}{\multirow{2}{*}{Code Summarization (\underline{Ja$\rightarrow$NL})}} & Ja & 57.86 & 39.92 & 42.50 & 63.95 & 44.01 & 43.50 & 17.95 (31.02\%) \\
 & \multicolumn{2}{c}{} & NL & 18.46 & 6.54 & 18.00 & 19.80 & 6.90 & 19.00 & 4.51 (26.27\%) \\ \cline{4-11}
 & \multicolumn{2}{c}{\multirow{2}{*}{Code Summarization (\underline{Py$\rightarrow$NL})}} & Py & 71.41 & 50.98 & 56.00 & 71.55 & 50.81 & 57.00 & 21.48 (30.08\%) \\
 & \multicolumn{2}{c}{} & NL & 16.60 & 5.64 & 16.50 & 17.79 & 5.94 & 17.50 & 4.11 (26.41\%) \\ \hline
\multirow{8}{*}{StarCoder} & \multicolumn{2}{c}{\multirow{2}{*}{Code Generation (\underline{NL$\rightarrow$Ja})}} & NL & 28.71 & 13.35 & 26.00 & 29.51 & 13.52 & 26.50 & 6.33 (22.05\%) \\
 & \multicolumn{2}{c}{} & Ja & 45.02 & 22.99 & 38.00 & 45.90 & 23.28 & 39.00 & 10.46 (23.23\%) \\ \cline{4-11}
 & \multicolumn{2}{c}{\multirow{2}{*}{Code Generation (\underline{NL$\rightarrow$Py})}} & NL & 36.21 & 36.57 & 23.00 & 37.98 & 28.77 & 23.50 & 6.05 (19.08\%) \\
 & \multicolumn{2}{c}{} & Py & 77.21 & 47.64 & 69.00 & 76.85 & 52.73 & 69.50 & 25.68 (33.26\%) \\ \cline{4-11}
 & \multicolumn{2}{c}{\multirow{2}{*}{Code Summarization (\underline{Ja$\rightarrow$NL})}} & Ja & 59.94 & 36.20 & 53.50 & 62.15 & 36.92 & 53.50 & 13.03 (23.33\%) \\
 & \multicolumn{2}{c}{} & NL & 16.61 & 5.84 & 15.00 & 16.97 & 6.03 & 15.00 & 2.69 (17.31\%) \\ \cline{4-11}
 & \multicolumn{2}{c}{\multirow{2}{*}{Code Summarization (\underline{Py$\rightarrow$NL})}} & Py & 72.27 & 55.85 & 59.50 & 71.32 & 65.02 & 58.00 & 20.06 (30.77\%) \\
 & \multicolumn{2}{c}{} & NL & 12.59 & 3.93 & 12.00 & 13.61 & 4.32 & 13.00 & 3.97 (33.49\%) \\ \bottomrule
\end{tabular}
\end{table}

\subsection{Implementations}
\label{Implementations}
For pre-training of the PLMs, we perform 20 epochs with a fixed learning rate of $1 \times 10^{-5}$ and a batch size of 32. For fine-tuning on downstream tasks, we train each model for 50 epochs on all three tasks to ensure full convergence and select the model that achieves the best performance on the validation set. Our fine-tuning framework is adapted from the replication package of CodeXGLUE’s \cite{lu2021codexglue}. During inference, we fetch the first output for evaluation.
To enable direct inference by LLM, we deploy LLaMA and StarCoder locally with the model initialization and weight loading from Hugging Face \cite{wolf2020transformers}. 
Subsequently, we perform inference on code tasks using a series of basic prompts, and their specific examples can be found at \cite{anonymous:online}, due to the page limit.
To alleviate the chaos of outputs, we additionally provide a one-shot example for in-context learning and retrieve only the first output for evaluation.
The inference process is conducted with the following decoding parameters: \textit{temperature} = 0.1, \textit{top\_p} = 0.95, and \textit{top\_k} = 50.

To mitigate the influence of randomness and enhance the reliability of the results, all experiments concerning PLMs and LLMs are repeated five times, respectively.
Besides, we employ the non-parametric Mann–Whitney U test~\cite{mann1947test} with a two-sided hypothesis to assess whether the performance differences between experimental and control groups are statistically significant. For each test, the null hypothesis is defined as: ``There is no statistically significant difference between the results of the two groups.'' All statistical tests are conducted at a significance level of $\alpha$ = 0.05, corresponding to a 95\% confidence level for determining statistical significance.  
All experiments above are conducted on six Nvidia A40 GPUs with 48GB of memory for each. The pre-training, fine-tuning, and inference of two PLMs across three code tasks and two PLs require 3,500 GPU-hours, while the inference of two LLMs for all experiments requires 100 GPU-hours.

\section{Experimental Results}
\label{Experimental Result}
This section discusses the experimental results targeting each RQ proposed in Section \ref{Research Questions}. 

\begin{table*}[htbp]
  \centering
  \setlength{\abovecaptionskip}{0cm}
  \setlength{\tabcolsep}{3.5pt}
  \begin{threeparttable}
    \caption{Results of Input-Only Contamination}
    \label{tab:1}
    \scriptsize

    \begin{tabular}{@{} l c c c c @{}}
      \toprule
      \multirow{2.5}{*}{\textbf{Metric}} & 
      \multicolumn{4}{c}{\textbf{Model Performance} (Experimental Group / Control Group / \textit{p}-value)} \\
      \cmidrule(l){2-5}
      & \textbf{RoBERTa} & \textbf{GPT-2} & \textbf{LLaMA} & \textbf{StarCoder} \\
      \midrule

      \multicolumn{5}{c}{\textit{Code Translation: Java $\rightarrow$ C\#}} \\

      Cry-BLEU &
      $75.57_{\pm0.93}$ / \textbf{76.00}$_{\pm0.18}$ / 0.579 &
      \textbf{31.18}$_{\pm0.56}$ / $30.88_{\pm0.10}$ / 0.500 &
      $46.06_{\pm1.60}$ / \textbf{45.12}$_{\pm1.42}$ / 0.155 &
      $52.03_{\pm1.66}$ / \textbf{54.17}$_{\pm0.49}$ / 0.952 \\

      CodeBLEU &
      $84.18_{\pm0.90}$ / \textbf{84.20}$_{\pm0.19}$ / 0.111 &
      33.91$_{\pm0.24}$ / $\textbf{33.97}_{\pm0.14}$ / 0.655 &
      $\textbf{51.60}_{\pm0.46}$ / 51.15$_{\pm0.72}$ / 0.274 &
      $\textbf{51.73}_{\pm0.30}$ / 51.44$_{\pm0.23}$ / 0.087 \\

      \addlinespace[2pt]
      \multicolumn{5}{c}{\textit{Code Translation: Python $\rightarrow$ Java}} \\
      \addlinespace[2pt]

    Cry-BLEU &
    \textbf{55.64}$_{\pm0.41}$ / $55.38_{\pm0.35}$ / 0.274 &
    $14.02_{\pm0.06}$ / \textbf{14.29}$_{\pm0.13}$ / 0.997 &
    \textbf{21.48}$_{\pm0.21}$ / $21.45_{\pm0.14}$ / 0.500 &
    \textbf{25.84}$_{\pm0.47}$ / $25.26_{\pm0.55}$ / 0.111 \\

    CodeBLEU &
    \textbf{58.24}$_{\pm0.26}$ / $58.06_{\pm0.15}$ / 0.155 &
    \textbf{33.08}$_{\pm0.08}$ / $33.01_{\pm0.12}$ / 0.229 &
    \textbf{25.26}$_{\pm0.06}$ / $25.18_{\pm0.10}$ / 0.170 &
    \textbf{30.29}$_{\pm0.42}$ / $29.83_{\pm0.27}$ / 0.111 \\
    \midrule
      \multicolumn{5}{c}{\textit{Code Generation: NL $\rightarrow$ Java}} \\

    Cry-BLEU &
    $13.21_{\pm0.19}$ / \textbf{13.24}$_{\pm0.30}$ / 0.799 &
    $1.02_{\pm0.12}$ / \textbf{1.04}$_{\pm0.14}$ / 0.910 &
    —— / —— / —— &
    —— / —— / —— \\

    CodeBLEU &
    \textbf{36.65}$_{\pm0.84}$ / $36.15_{\pm0.89}$ / 0.210 &
    $7.21_{\pm0.24}$ / \textbf{7.47}$_{\pm0.38}$ / 0.296 &
    —— / —— / —— &
    —— / —— / —— \\

      \addlinespace[2pt]
      \multicolumn{5}{c}{\textit{Code Generation: NL $\rightarrow$ Python}} \\
      \addlinespace[2pt]

    Cry-BLEU &
    \textbf{27.62}$_{\pm0.27}$ / $27.56_{\pm0.20}$ / 0.500 &
    \textbf{1.62}$_{\pm0.17}$ / $1.56_{\pm0.20}$ / 0.500 &
    —— / —— / —— &
    —— / —— / —— \\

    CodeBLEU &
    $30.41_{\pm0.19}$ / \textbf{30.46}$_{\pm0.15}$ / 0.700 &
    $9.43_{\pm0.21}$ / \textbf{9.47}$_{\pm0.15}$ / 0.579 &
    —— / —— / —— &
    —— / —— / —— \\
    \midrule
      \multicolumn{5}{c}{\textit{Code Summarization: Java $\rightarrow$ NL}} \\

      BLEU &
      \textbf{39.24}$_{\pm0.28}$ / 38.68$_{\pm0.42}$ / 0.296 &
      $2.77_{\pm0.04}$  / \textbf{2.80}$_{\pm0.06}$  / 0.249 &
      $10.60_{\pm0.67}$ / \textbf{10.95}$_{\pm0.92}$ / 0.531 &
      \textbf{9.99}$_{\pm0.48}$ / $9.73_{\pm1.09}$ / 0.531 \\

      ROUGE-L &
    \textbf{51.40}$_{\pm0.29}$ / $51.16_{\pm0.38}$ / 0.147 &
    \textbf{5.60}$_{\pm0.01}$ / $5.50_{\pm0.29}$ / 0.953 &
    $36.44_{\pm0.98}$ / \textbf{37.09}$_{\pm0.48}$ / 0.875 &
    \textbf{36.67}$_{\pm1.02}$ / $36.13_{\pm0.67}$ / 0.210 \\

      \addlinespace[2pt]
      \multicolumn{5}{c}{\textit{Code Summarization: Python $\rightarrow$ NL}} \\
      \addlinespace[2pt]

      BLEU &
      \textbf{37.39}$_{\pm0.21}$ / 37.09$_{\pm0.25}$ / 0.144 &
      $7.12_{\pm0.05}$ / \textbf{7.15}$_{\pm0.03}$ / 0.401 &
      \textbf{10.17}$_{\pm0.43}$ / 9.82$_{\pm0.26}$ / 0.296 &
      \textbf{6.98}$_{\pm0.33}$ / 6.75$_{\pm0.21}$ / 0.296 \\

      ROUGE-L &
    \textbf{48.50}$_{\pm0.36}$ / $48.29_{\pm0.20}$ / 0.155 &
    $15.76_{\pm0.04}$ / \textbf{15.78}$_{\pm0.04}$ / 0.771 &
    \textbf{38.64}$_{\pm0.45}$ / $38.29_{\pm0.35}$ / 0.173 &
    \textbf{34.45}$_{\pm0.17}$ / $34.02_{\pm0.42}$ / 0.075 \\

      \bottomrule
    \end{tabular}

    \begin{tablenotes}
      \footnotesize
      \item[] \textit{Note}: 
      Owing to the page limit, each cell only showcases results with their average score and standard deviation of experimental and control groups, respectively, as well as corresponding statistical tests (\textit{p}-value). In addition, \textit{p}-values less than 0.05 are indicated in \textcolor{red}{red}. The “Cry-BLEU” metric reported in the table refers to CrystalBLEU.
      The subsequent tables follow the same convention.
    \end{tablenotes}
  \end{threeparttable}
\end{table*}

\subsection{Results of the Input-Only Contamination Setting (RQ1)} \label{Results of Input-Only Contamination Setting (RQ1)}
As illustrated in Section \ref{RQ1: Does Input-Only Contamination Lead to Performance Overestimation?}, we conduct experiments on code translation, generation, and summarization tasks for PLMs while carrying out code translation and summarization tasks for LLMs. Below, we introduce their experimental results and corresponding analysis, respectively. 

\textbf{As for PLMs (Pre-train, fine-tune, then infer)}, the performance of RoBERTa and GPT-2 under control and experimental groups is reported in Table~\ref{tab:1}. 
Overall, the contaminated experimental groups show only marginal and inconsistent changes compared with the uncontaminated control groups across all three tasks and the reported metrics (Cry-BLEU/CodeBLEU for code translation and generation, BLEU/ROUGE-L for code summarization). For RoBERTa, the largest relative gain appears on code summarization (e.g., Java$\rightarrow$NL BLEU: +1.45\%), while other settings remain similarly small (e.g., NL$\rightarrow$Java CodeBLEU: +1.38\%). For GPT-2, the relative gaps are also small and can be either positive or negative depending on the setting (e.g., Java$\rightarrow$C\# Cry-BLEU: +0.97\%; Python$\rightarrow$Java Cry-BLEU: $-1.89$\%), and none of the comparisons is statistically significant (\textit{p}-values$>$0.05). These results suggest that the performance of PLMs will not be overestimated in the downstream tasks, even though their pre-training corpora are contaminated by the input portions of testing sets. This conclusion aligns with our expectation and is reasonable for explanation, as PLMs do not learn any mapping from input to output portions of the testing sets. Their pre-training corpora only contain questions, but no answers.

\textbf{As for LLMs (Pre-train, then infer)}, their experimental results of the input-only contamination setting are also listed in Table \ref{tab:1}.
As can be seen, the inference performance of LLaMA and StarCoder in the contaminated experimental group and the uncontaminated control group still has no obvious discrepancy among the five consecutive trials, manifesting a relative performance difference of only (-3.95\%)--3.41\% across both code translation and summarization tasks. Besides, statistical tests further verify the nonsignificance of this difference with \textit{p}-values all larger than 0.05. The reason is the same as that we mentioned in the PLM part, as LLMs only saw the input portions during their pre-training, considering the substantial discrepancies between inputs and outputs in code tasks, by no means can they carry any superiority in generating the output portions. 

\newenvironment{boxK}{\begin{tcolorbox}[
    colback=gray!10,
    colframe=black!80,
    left=2pt,right=2pt,top=1pt,bottom=1pt, 
]}{\end{tcolorbox}}
\begin{boxK}
\textbf{Finding 1:} Input-only contamination during the pre-training stage does not lead to significant performance overestimation in downstream tasks for either PLMs or LLMs, fluctuating at (-3.95\%)--3.41\% against the uncontaminated control group at most (\textit{p}-values$>$0.05).

\end{boxK}

\subsection{Results of the Output-only Contamination Setting (RQ2)}
As elaborated in Section \ref{RQ2: Does Output-Only Contamination Lead to Performance Overestimation?}, PLMs are still experimented on all studied code tasks, while LLMs are assessed on two of the code tasks, i.e., code translation and generation. 
\textbf{As for PLMs (Pre-train, fine-tune, then infer)}, the results for RoBERTa and GPT-2 under this setting are shown in Table~\ref{tab:2}. Similar to the input-only contamination experiments, there is no significant difference in performance between the contaminated experimental group and the uncontaminated control group, regardless of any code task and PLMs under test. Concretely, RoBERTa shows small gains on most code translation and generation settings under output-only contamination (e.g., Java$\rightarrow$C\# CodeBLEU: +0.34\%; NL$\rightarrow$Python Cry-BLEU: +0.98\%), while GPT-2 presents mixed outcomes and can underperform on some metrics (e.g., Java$\rightarrow$NL ROUGE-L: $-6.36$\%). However, the above differences are still minimal, and the statistical tests among their corresponding five trials do not manifest any significance with \textit{p}-values all larger than 0.05.
This indicates that exposure to the output portions of the testing set during the pre-training stage does not cause significant performance overestimation for PLMs.


\textbf{As for LLMs (Pre-train, then infer)}, the inference results under the output-only contamination setting are presented in Table~\ref{tab:2} as well. As can be seen, both LLMs do not show consistent superiority under output-only contamination: while there are only minor changes on code translation (e.g., LLaMA Cry-BLEU on Java$\rightarrow$C\#: +0.34\%), the contaminated groups generally underperform on code generation for both models (e.g., LLaMA CodeBLEU on NL$\rightarrow$Java: $-4.51$\%; StarCoder Cry-BLEU on NL$\rightarrow$Java: $-5.87$\%). 
In addition, none of the statistical tests demonstrate significant differences with \textit{p}-values larger than 0.05 consistently. This suggests that contamination appearing on output portions of testing sets will not induce significant performance overestimation in downstream tasks.
As we explained in Section \ref{Results of Input-Only Contamination Setting (RQ1)}, models only saw the partial samples of testing sets, which is not helpful for them to infer the rest of the samples, as the differences between input and output portions of code tasks are substantial.

\begin{boxK}
\textbf{Finding 2:} Both PLMs and LLMs show no significant performance overestimation on downstream tasks when contaminating the pre-training corpora with their test sets' output portions. The performance fluctuation between experimental and control groups is restricted in (-9.19\%)--1.29\% at most (\textit{p}-values$>$0.05).
\end{boxK}

\begin{table*}[htbp] 

  \centering 
  \setlength{\abovecaptionskip}{0cm}
  \setlength{\tabcolsep}{3.5pt} 
    \caption{Results of Output-Only Contamination} 
    \label{tab:2} 
    \scriptsize 

    \begin{tabular}{@{} l c c c c @{}} 
      \toprule 
      \multirow{2.5}{*}{\textbf{Metric}} & 
      \multicolumn{4}{c}{\textbf{Model Performance} (Experimental Group / Control Group / \textit{p}-value)} \\ 
      \cmidrule(l){2-5} 
      & \textbf{RoBERTa} & \textbf{GPT-2} & \textbf{LLaMA} & \textbf{StarCoder} \\ 
      \midrule 

      \multicolumn{5}{c}{\textit{Code Translation: Java $\rightarrow$ C\#}} \\ 
      Cry-BLEU &
    \textbf{76.17}$_{\pm0.14}$ / $76.00_{\pm0.18}$ / 0.087 &
    $30.81_{\pm0.11}$ / \textbf{30.88}$_{\pm0.10}$ / 0.737 &
    \textbf{44.51}$_{\pm0.21}$ / $44.36_{\pm0.14}$ / 0.111 &
    \textbf{45.08}$_{\pm0.61}$ / $44.93_{\pm0.98}$ / 0.095 \\ 
      CodeBLEU &
    \textbf{84.49}$_{\pm0.28}$ / $84.20_{\pm0.19}$ / 0.075 &
    $33.67_{\pm0.62}$ / \textbf{33.97}$_{\pm0.34}$ / 0.996 &
    \textbf{44.93}$_{\pm0.19}$ / $44.36_{\pm0.58}$ / 0.201 &
    $53.32_{\pm0.36}$ / \textbf{54.45}$_{\pm0.77}$ / 0.992 \\ 

      \addlinespace[2pt] 
      \multicolumn{5}{c}{\textit{Code Translation: Python $\rightarrow$ Java}} \\ 
      \addlinespace[2pt] 

      Cry-BLEU &
    \textbf{55.43}$_{\pm0.23}$ / $55.38_{\pm0.35}$ / 0.542 &
    $14.13_{\pm0.12}$ / \textbf{14.29}$_{\pm0.13}$ / 0.942 &
    \textbf{37.96}$_{\pm0.41}$ / $37.57_{\pm0.32}$ / 0.111 &
    $46.38_{\pm0.75}$ / \textbf{46.56}$_{\pm0.38}$ / 0.726 \\
      CodeBLEU &
    \textbf{58.19}$_{\pm0.17}$ / $58.06_{\pm0.15}$ / 0.173 &
    \textbf{33.03}$_{\pm0.09}$ / $33.01_{\pm0.12}$ / 0.542 &
    $42.39_{\pm0.44}$ / \textbf{42.74}$_{\pm0.68}$ / 0.845 &
    $49.28_{\pm0.50}$ / \textbf{50.88}$_{\pm0.38}$ / 1.000 \\ 
      \midrule 

      \multicolumn{5}{c}{\textit{Code Generation: NL $\rightarrow$ Java}} \\ 
      Cry-BLEU &
    \textbf{13.28}$_{\pm0.80}$ / $13.24_{\pm0.30}$ / 0.336 &
    $1.04_{\pm0.03}$ / \textbf{1.04}$_{\pm0.04}$ / 0.586 &
    $44.77_{\pm1.10}$ / \textbf{48.79}$_{\pm1.81}$ / 1.000 &
    $47.83_{\pm2.03}$ / \textbf{50.81}$_{\pm2.04}$ / 0.972 \\ 
      CodeBLEU &
    \textbf{36.26}$_{\pm2.81}$ / $36.15_{\pm0.89}$ / 0.155 &
    $7.36_{\pm0.11}$ / \textbf{7.47}$_{\pm0.18}$ / 0.845 &
    $40.70_{\pm1.19}$ / \textbf{42.62}$_{\pm1.24}$ / 0.972 &
    $48.61_{\pm1.54}$ / \textbf{51.07}$_{\pm0.93}$ / 0.992 \\

      \addlinespace[2pt] 
      \multicolumn{5}{c}{\textit{Code Generation: NL $\rightarrow$ Python}} \\ 
      \addlinespace[2pt] 

      Cry-BLEU &
    \textbf{27.83}$_{\pm0.23}$ / $27.56_{\pm0.20}$ / 0.104 &
    \textbf{1.51}$_{\pm0.34}$ / $1.56_{\pm0.29}$ / 0.155 &
    $24.08_{\pm0.58}$ / \textbf{24.70}$_{\pm0.80}$ / 0.889 &
    $39.82_{\pm0.23}$ / \textbf{40.63}$_{\pm0.33}$ / 1.000 \\ 
      CodeBLEU &
    \textbf{30.59}$_{\pm0.10}$ / $30.46_{\pm0.15}$ / 0.111 &
    \textbf{9.61}$_{\pm0.09}$ / $9.47_{\pm0.15}$ / 0.125 &
    $21.65_{\pm0.59}$ / \textbf{23.84}$_{\pm0.37}$ / 1.000 &
    $37.77_{\pm0.24}$ / \textbf{38.20}$_{\pm0.56}$ / 0.925 \\ 
      \midrule 

      \multicolumn{5}{c}{\textit{Code Summarization: Java $\rightarrow$ NL}} \\ 
      BLEU & 
      \textbf{39.04}$_{\pm0.66}$ / 38.68$_{\pm0.42}$ / 1.000 & 
      $2.77_{\pm0.06}$ / \textbf{2.80}$_{\pm0.06}$ / 0.528 & 
      —— / —— / —— & 
      —— / —— / —— \\ 
      ROUGE-L &
    \textbf{51.54}$_{\pm0.55}$ / $51.16_{\pm0.38}$ / 0.210 &
    $5.15_{\pm0.08}$ / \textbf{5.50}$_{\pm0.29}$ / 0.952 &
    —— / —— / —— &
    —— / —— / —— \\ 

      \addlinespace[2pt] 
      \multicolumn{5}{c}{\textit{Code Summarization: Python $\rightarrow$ NL}} \\ 
      \addlinespace[2pt] 

      BLEU & 
      $36.99_{\pm0.14}$ / \textbf{37.09}$_{\pm0.25}$ / 0.676 & 
      $7.13_{\pm0.03}$ / \textbf{7.15}$_{\pm0.03}$ / 0.399 & 
      —— / —— / —— & 
      —— / —— / —— \\ 
      ROUGE-L &
    \textbf{48.29}$_{\pm0.28}$ / $48.29_{\pm0.20}$ / 0.542 &
    $15.73_{\pm0.10}$ / \textbf{15.78}$_{\pm0.04}$ / 0.896 &
    —— / —— / —— &
    —— / —— / —— \\ 

      \bottomrule 
    \end{tabular} 
\end{table*}

\begin{table*}[htbp] 
  \centering 
  \setlength{\abovecaptionskip}{0cm}
  \setlength{\tabcolsep}{3.5pt}
  \begin{threeparttable} 
    \caption{Results of Unpaired Contamination} 
    \label{tab:3} 
    \scriptsize 

    \begin{tabular}{@{} l c c c c @{}} 
      \toprule 
      \multirow{2.5}{*}{\textbf{Metric}} & 
      \multicolumn{4}{c}{\textbf{Model Performance} (Experimental Group / Control Group / \textit{p}-value)} \\ 
      \cmidrule(l){2-5} 
      & \textbf{RoBERTa} & \textbf{GPT-2} & \textbf{LLaMA} & \textbf{StarCoder} \\ 
      \midrule 

      \multicolumn{5}{c}{\textit{Code Translation: Java $\rightarrow$ C\#}} \\ 
      Cry-BLEU &
    \textbf{76.16}$_{\pm0.16}$ / $76.00_{\pm0.18}$ / 0.111 &
    \textbf{30.95}$_{\pm0.17}$ / $30.88_{\pm0.80}$ / 0.147 &
    \textbf{38.27}$_{\pm1.71}$ / $36.57_{\pm2.06}$ / 0.075 &
    $51.02_{\pm0.49}$ / \textbf{51.76}$_{\pm0.53}$ / 0.972 \\ 
      CodeBLEU &
    \textbf{84.35}$_{\pm0.17}$ / $84.20_{\pm0.19}$ / 0.172 &
    \textbf{34.00}$_{\pm0.61}$ / $33.97_{\pm0.14}$ / 0.376 &
    $46.20_{\pm5.06}$ / \textbf{50.00}$_{\pm0.68}$ / 0.726 &
    $54.64_{\pm1.27}$ / \textbf{55.79}$_{\pm0.95}$ / 0.925 \\ 

      \addlinespace[2pt] 
      \multicolumn{5}{c}{\textit{Code Translation: Python $\rightarrow$ Java}} \\ 
      \addlinespace[2pt] 

      Cry-BLEU &
    \textbf{55.99}$_{\pm0.60}$ / $55.38_{\pm0.35}$ / 0.075 &
    $14.12_{\pm0.07}$ / \textbf{14.29}$_{\pm0.13}$ / 0.984 &
    \textbf{7.15}$_{\pm0.18}$ / $7.17_{\pm0.61}$ / 0.104 &
    \textbf{12.16}$_{\pm1.58}$ / $11.97_{\pm0.75}$ / 0.173 \\ 
      CodeBLEU &
    \textbf{58.40}$_{\pm0.30}$ / $58.06_{\pm0.45}$ / 0.210 &
    \textbf{33.04}$_{\pm0.05}$ / $33.01_{\pm0.12}$ / 0.500 &
    $24.80_{\pm0.05}$ / \textbf{26.54}$_{\pm0.07}$ / 1.000 &
    $32.57_{\pm0.16}$ / \textbf{33.52}$_{\pm0.25}$ / 1.000 \\ 
      \midrule 

      \multicolumn{5}{c}{\textit{Code Generation: NL $\rightarrow$ Java}} \\ 
      Cry-BLEU &
    \textbf{13.36}$_{\pm0.32}$ / $13.24_{\pm0.30}$ / 0.500 &
    $1.02_{\pm0.03}$ / \textbf{1.04}$_{\pm0.04}$ / 0.815 &
    —— / —— / —— &
    —— / —— / —— \\
      CodeBLEU &
    $35.81_{\pm1.06}$ / \textbf{36.15}$_{\pm0.89}$ / 0.735 &
    $7.12_{\pm0.05}$ / \textbf{7.47}$_{\pm0.18}$ / 1.000 &
    —— / —— / —— &
    —— / —— / —— \\ 

      \addlinespace[2pt] 
      \multicolumn{5}{c}{\textit{Code Generation: NL $\rightarrow$ Python}} \\ 
      \addlinespace[2pt] 

      Cry-BLEU &
    \textbf{27.77}$_{\pm0.19}$ / $27.56_{\pm0.20}$ / 0.145 &
    \textbf{1.62}$_{\pm0.23}$ / $1.56_{\pm0.20}$ / 0.173 &
    —— / —— / —— &
    —— / —— / —— \\
      CodeBLEU &
    \textbf{30.72}$_{\pm0.31}$ / $30.46_{\pm0.15}$ / 0.075 &
    \textbf{9.69}$_{\pm0.30}$ / $9.47_{\pm0.15}$ / 0.125 &
    —— / —— / —— &
    —— / —— / —— \\ 
      \midrule 

      \multicolumn{5}{c}{\textit{Code Summarization: Java $\rightarrow$ NL}} \\ 
      BLEU & 
      $38.61_{\pm0.58}$ / \textbf{38.68}$_{\pm0.42}$ / 0.296 & 
      \textbf{2.81}$_{\pm0.03}$ / 2.80$_{\pm0.06}$ / 1.000 & 
      —— / —— / —— & 
      —— / —— / —— \\ 
      ROUGE-L &
    $50.84_{\pm0.36}$ / \textbf{51.16}$_{\pm0.38}$ / 0.827 &
    \textbf{5.69}$_{\pm0.05}$ / $5.50_{\pm0.29}$ / 0.087 &
    —— / —— / —— &
    —— / —— / —— \\ 

      \addlinespace[2pt] 
      \multicolumn{5}{c}{\textit{Code Summarization: Python $\rightarrow$ NL}} \\ 
      \addlinespace[2pt] 

      BLEU & 
      $37.06_{\pm0.45}$ / \textbf{37.09}$_{\pm0.25}$ / 0.835 & 
      $7.12_{\pm0.06}$ / \textbf{7.15}$_{\pm0.03}$ / 0.142 & 
      —— / —— / —— & 
      —— / —— / —— \\ 
      ROUGE-L &
    $48.04_{\pm0.38}$ / \textbf{48.29}$_{\pm0.20}$ / 0.889 &
    $15.71_{\pm0.09}$ / \textbf{15.78}$_{\pm0.04}$ / 0.896 &
    —— / —— / —— &
    —— / —— / —— \\ 
      \bottomrule 
    \end{tabular} 
  \end{threeparttable} 
\end{table*}

\subsection{Results of the Unpaired Contamination Setting (RQ3)}
As presented in Section \ref{RQ3: Does Unpaired Contamination Lead to Performance Overestimation?}, unpaired contamination experiments are conducted on all code tasks for PLMs, while LLMs are only evaluated on the code translation task solely. 

Similarly, we first present the results of \textbf{PLMs (Pre-train, fine-tune, then infer)}, which are shown in Table~\ref{tab:3}. 
As can be seen, the performance gaps between contaminated experimental groups and uncontaminated control groups are consistently small and do not exhibit a systematic advantage under unpaired contamination. On code translation and generation, RoBERTa shows only marginal gains on Cry-BLEU in several settings (e.g., Python$\rightarrow$Java Cry-BLEU: +1.10\%; NL$\rightarrow$Python Cry-BLEU: +0.76\%), while its CodeBLEU can be slightly higher or lower depending on the task (e.g., NL$\rightarrow$Java CodeBLEU: $-0.94$\%). GPT-2 exhibits similarly mixed outcomes with small differences (e.g., NL$\rightarrow$Python CodeBLEU: +2.32\%). For code summarization, the contaminated groups do not outperform consistently either (e.g., RoBERTa Java$\rightarrow$NL ROUGE-L: $-0.63$\%). Moreover, the statistical tests demonstrate that there is no significant difference between the two groups above with \textit{p}-value larger than 0.05. 
This suggests that merely observing both inputs and outputs independently is insufficient for the PLMs to learn their correspondence in downstream tasks, as the input-output mappings are not constructed during the pre-training stage.

The experimental results of the unpaired contamination setting for \textbf{LLMs (Pre-train, then infer)} are shown in Table~\ref{tab:3}. As can be seen, neither the contaminated experimental group nor the uncontaminated control group demonstrates significant superiority over the other on the code translation task. Specifically, Cry-BLEU shows small and mixed changes (e.g., on Java$\rightarrow$C\#, LLaMA: +4.65\% while StarCoder: $-1.43$\%), and CodeBLEU also varies without a consistent trend (e.g., LLaMA on Java$\rightarrow$C\#: $-7.60$\%). 
Meanwhile, the above performance differences still cannot be considered significant, as the \textit{p}-values of the statistical tests are larger than 0.05. 
The above experimental results are rational because, theoretically, unpaired contamination only involves pre-training with both input and output portions of testing sets, but the mapping between them is unlearned. 

\begin{boxK}
\textbf{Finding 3:} Unpaired contamination does not lead to significant performance overestimation in downstream tasks for either PLMs or LLMs, fluctuating at (-7.60\%)--4.65\% against the uncontaminated control group at most (\textit{p}-values$>$0.05).
\end{boxK}

\begin{table*}[htbp] 
  \centering 
  \setlength{\abovecaptionskip}{0cm}
  \setlength{\tabcolsep}{3.5pt} 
    \caption{Results of Paired Contamination} 
    \label{tab:4} 
    \scriptsize 
    \begin{tabular}{@{} l c c c c @{}} 
      \toprule 
      \multirow{2.5}{*}{\textbf{Metric}} & 
      \multicolumn{4}{c}{\textbf{Model Performance} (Experimental Group / Control Group / \textit{p}-value)} \\ 
      \cmidrule(l){2-5} 
      & \textbf{RoBERTa} & \textbf{GPT-2} & \textbf{LLaMA} & \textbf{StarCoder} \\ 
      \midrule 

      \multicolumn{5}{c}{\textit{Code Translation: Java $\rightarrow$ C\#}} \\ 
      Cry-BLEU &
    \textbf{76.09}$_{\pm0.26}$ / $76.00_{\pm0.18}$ / 0.345 &
    \textbf{30.94}$_{\pm0.02}$ / $30.88_{\pm0.10}$ / 0.300 &
    —— / —— / —— &
    —— / —— / —— \\ 
    CodeBLEU &
    \textbf{84.40}$_{\pm0.25}$ / $84.20_{\pm0.19}$ / 0.111 &
    \textbf{34.04}$_{\pm0.09}$ / $33.97_{\pm0.14}$ / 0.200 &
    —— / —— / —— &
    —— / —— / —— \\ 

      \addlinespace[2pt] 
      \multicolumn{5}{c}{\textit{Code Translation: Python $\rightarrow$ Java}} \\ 
      \addlinespace[2pt] 

      Cry-BLEU &
    \textbf{55.78}$_{\pm0.56}$ / $55.38_{\pm0.35}$ / 0.155 &
    $14.09_{\pm0.10}$ / \textbf{14.29}$_{\pm0.13}$ / 0.992 &
    —— / —— / —— &
    —— / —— / —— \\ 
     CodeBLEU &
    \textbf{58.29}$_{\pm0.34}$ / $58.06_{\pm0.15}$ / 0.232 &
    \textbf{33.11}$_{\pm0.11}$ / $33.01_{\pm0.12}$ / 0.071 &
    —— / —— / —— &
    —— / —— / —— \\
      \midrule 

      \multicolumn{5}{c}{\textit{Code Generation: NL $\rightarrow$ Java}} \\ 
      Cry-BLEU &
$13.24_{\pm0.38}$ / \textbf{13.24}$_{\pm0.30}$ / 0.500 &
\textbf{1.32}$_{\pm0.09}$ / $1.04_{\pm0.64}$ / 0.087 &
\textbf{21.33}$_{\pm0.92}$ / $19.49_{\pm1.82}$ / \textcolor{red}{0.048} &
\textbf{27.90}$_{\pm1.11}$ / $25.35_{\pm1.90}$ / \textcolor{red}{0.048} \\ 
      CodeBLEU &
\textbf{37.35}$_{\pm1.38}$ / $36.15_{\pm0.89}$ / 0.111 &
\textbf{7.52}$_{\pm0.21}$ / $7.47_{\pm0.18}$ / 0.155 &
\textbf{22.08}$_{\pm0.95}$ / $20.10_{\pm1.65}$ / \textcolor{red}{0.048} &
\textbf{27.06}$_{\pm0.71}$ / $25.20_{\pm1.55}$ / \textcolor{red}{0.048} \\ 

      \addlinespace[2pt] 
      \multicolumn{5}{c}{\textit{Code Generation: NL $\rightarrow$ Python}} \\ 
      \addlinespace[2pt] 

      Cry-BLEU &
\textbf{27.87}$_{\pm0.24}$ / $27.56_{\pm0.20}$ / 0.071 &
\textbf{1.67}$_{\pm0.24}$ / $1.56_{\pm0.20}$ / 0.071 &
\textbf{6.98}$_{\pm0.25}$ / $5.98_{\pm0.23}$ / \textcolor{red}{0.004} &
\textbf{5.75}$_{\pm0.46}$ / $4.72_{\pm0.18}$ / \textcolor{red}{0.004} \\ 
      CodeBLEU &
\textbf{30.61}$_{\pm0.14}$ / $30.46_{\pm0.15}$ / 0.087 &
\textbf{9.67}$_{\pm0.14}$ / $9.47_{\pm0.15}$ / 0.057 &
\textbf{12.78}$_{\pm0.41}$ / $12.10_{\pm0.71}$ / \textcolor{red}{0.048} &
\textbf{10.56}$_{\pm0.35}$ / $9.32_{\pm0.18}$ / \textcolor{red}{0.006} \\ 
      \midrule 

      \multicolumn{5}{c}{\textit{Code Summarization: Java $\rightarrow$ NL}} \\ 
      BLEU & 
      \textbf{39.06}$_{\pm0.60}$ / 38.68$_{\pm0.42}$ / 0.676 & 
      $2.80_{\pm0.05}$ / \textbf{2.80}$_{\pm0.06}$ / 0.834 & 
      \textbf{10.11}$_{\pm0.11}$ / 8.77$_{\pm0.28}$ / \textcolor{red}{0.012} & 
      \textbf{11.60}$_{\pm0.30}$ / $9.92_{\pm0.16}$ / \textcolor{red}{0.012} \\ 
      ROUGE-L &
$51.13_{\pm0.39}$ / \textbf{51.16}$_{\pm0.38}$ / 0.421 &
\textbf{5.76}$_{\pm0.19}$ / $5.50_{\pm0.29}$ / 0.210 &
\textbf{45.32}$_{\pm0.32}$ / $39.56_{\pm0.35}$ / \textcolor{red}{0.006} &
\textbf{47.87}$_{\pm0.31}$ / $40.42_{\pm0.36}$ / \textcolor{red}{0.004} \\

      \addlinespace[2pt] 
      \multicolumn{5}{c}{\textit{Code Summarization: Python $\rightarrow$ NL}} \\ 
      \addlinespace[2pt] 

      BLEU & 
      $37.51_{\pm0.36}$ / \textbf{37.09}$_{\pm0.25}$ / 0.095 & 
      7.17$_{\pm0.03}$ / 7.15$_{\pm0.03}$ / 0.598 & 
      \textbf{7.50}$_{\pm0.28}$ / 6.55$_{\pm0.15}$ / \textcolor{red}{0.012} & 
      $\textbf{7.21}_{\pm0.18}$ / 6.41$_{\pm0.13}$ / \textcolor{red}{0.012} \\ 
      ROUGE-L &
\textbf{48.44}$_{\pm0.26}$ / $48.29_{\pm0.20}$ / 0.232 &
$15.61_{\pm0.10}$ / \textbf{15.78}$_{\pm0.04}$ / 0.989 &
\textbf{37.20}$_{\pm0.21}$ / $34.93_{\pm0.61}$ / \textcolor{red}{0.006} &
\textbf{24.33}$_{\pm0.14}$ / $21.32_{\pm0.35}$ / \textcolor{red}{0.004} \\ 

      \bottomrule 
    \end{tabular} 
\end{table*}

\subsection{Results of the Paired Contamination Setting (RQ4)} \label{Results of the Paired Contamination Setting (RQ4)}
As illustrated in Section \ref{RQ4: Does Paired Contamination Lead to Performance Overestimation?}, paired contamination experiments are carried out on all three code tasks for PLMs, while for LLMs, only the code generation and summarization tasks are experimented with. 

Regarding the \textbf{PLMs (Pre-train, fine-tune, then infer)}, the experimental results are shown in Table~\ref{tab:4}. Overall, for the code translation task, experimental groups with data contamination slightly outperform the control group for both RoBERTa and GPT-2 on Cry-BLEU and CodeBLEU (e.g., RoBERTa Java$\rightarrow$C\# CodeBLEU: +0.24\%; GPT-2 Java$\rightarrow$C\# Cry-BLEU: +0.19\%). In contrast, on the code generation and summarization tasks, the two PLMs exhibit mixed performance trends across different metrics. For instance, RoBERTa shows modest gains on several settings (e.g., NL$\rightarrow$Java CodeBLEU: +3.32\%; Python$\rightarrow$NL BLEU: +1.13\%), while GPT-2 has both improvements and declines depending on the task/metric (e.g., Python$\rightarrow$Java Cry-BLEU: $-1.40$\%; Java$\rightarrow$NL ROUGE-L: +4.73\%). Even though they have diverse performance discrepancies between the two groups, the statistical tests still reflect non-significant differences for all the above results, with \textit{p}-values all larger than 0.05.
The limited contamination effect towards RoBERTa is reasonable, as its pre-training task is MLM, which is designed to learn to infill the masked tokens during the pre-training stage and is not aligned with the inference procedure of any downstream tasks. Besides, to implement the generative task, RoBERTa needs to be appended with a decoder structure, further weakening the impact of pre-training contamination on downstream tasks. On the contrary, the experimental results of GPT-2 are kind of surprising and opposite to our expectations, because it is an auto-regressive model~\cite{radford2018improving} and was pre-trained by the CLM task from scratch in our experiments. Considering that the input and output portions of the testing sets are contaminated in the pre-trained corpus of GPT-2, it should have learned how to generate correct tokens during the inference stage, theoretically. But the fact is not the case. We further explore possible explanations in Section \ref{Discussion}.

As for \textbf{LLMs (Pre-train, then infer)}, their experimental results are shown in Table~\ref{tab:4} as well. Apparently, both LLMs' performance in contaminated experimental groups significantly surpasses that in uncontaminated control groups. Specifically, on code generation, LLaMA achieves relative gains of 9.44\%--16.72\% on Cry-BLEU and 5.62\%--9.85\% on CodeBLEU, while StarCoder achieves 10.06\%--21.82\% on Cry-BLEU and 7.38\%--13.30\% on CodeBLEU. On code summarization, LLaMA gains 14.50\%--15.28\% on BLEU and 6.50\%--14.56\% on ROUGE-L, while StarCoder gains 12.48\%--16.94\% on BLEU and 14.12\%--18.43\% on ROUGE-L.
Moreover, statistical tests with \textit{p}-values smaller than 0.05 further testify to the significance of the difference. Considering that LLMs under test are all decoder-only architectures and are pre-trained with either CLM (i.e., LLaMA) or Fill-in-the-Middle (FIM) task (i.e., StarCoder), the explanation is straightforward. Because a decoder-only architecture implies its auto-regressive nature in practical usage. Besides, their pre-training tasks help them learn to generate subsequent tokens given prefixes, and even the FIM task broke away from the limitation of left-to-right generation \cite{bavarian2022efficient}. Both of the above reasons make LLMs' pre-training stage almost or exactly match their inference stage, thereby leading to performance overestimation of LLMs in downstream code intelligence tasks. 

\begin{boxK}
    \textbf{Finding 4:} Paired contamination does not significantly overestimate the performance of PLMs in downstream code tasks, with differences remaining non-significant across the reported metrics (\textit{p}-values$>$0.05). In contrast, LLMs exhibit statistically significant performance gains under paired contamination, improving by 5.62\%--21.82\% on code generation metrics (Cry-BLEU/CodeBLEU) and 6.50\%--18.43\% on code summarization metrics (BLEU/ROUGE-L) (\textit{p}-values$<$0.05).

\end{boxK}

\subsection{Discussion Experiments}
\label{Discussion}
As stated in RQ4, although we inject paired contaminated data into the pre-training stage of GPT-2, the overall model performance did not exhibit a significant improvement, even though it is an auto-regressive model and its pre-training task is almost the same as LLMs under test. The conclusion is completely opposite. Considering there is a fine-tuning stage between pre-training and inference for PLMs, we naturally propose our \textbf{RQ5: Whether the expected contamination effects appear if we drop the fine-tuning stage from experiments of GPT-2, and how fine-tuning influences the contamination effects?} 

To do this, we conduct a direct inference experiment using GPT-2 with previously pre-trained models in RQ4. The experimental results are shown in Table~\ref{tab:6}. We find that the model performance in the contaminated experimental group significantly outperforms that of the uncontaminated counterparts on all studied tasks. Concretely, on the code translation task, the contaminated group improves by 15.57\%--65.62\% in Cry-BLEU and 8.75\%--314.80\% in CodeBLEU. As for the code generation task, contamination brings gains of 14.29\%--85.47\% in Cry-BLEU and 3.85\%--152.17\% in CodeBLEU. Besides, GPT-2 also suffers an overestimation on the code summarization task, improving by 17.86\%--36.14\% in BLEU and 14.58\%--67.21\% in ROUGE-L. 
Besides, statistical tests also demonstrate the significance of the performance differences with \textit{p}-values all smaller than 0.05. This indicates that contamination indeed affects decoder-only PLMs (e.g., GPT-2), regardless of static- (e.g., Java) or dynamic-typed PLs (e.g., Python), making its performance overestimated with the pre-train-then-inference paradigm, but this effect was covered up by the subsequent fine-tuning. As mentioned in Section \ref{Datasets} and \ref{Implementations}, our fine-tuning was conducted with 100,000 training samples at most for each code task, and was trained for 50 epochs, respectively. Thus, we infer that fine-tuning brings about abundant underlying knowledge and task alignment specifications \cite{zhengspurious}, making the effects of contamination in the pre-training stage negligible. 

\begin{figure*}[!t]
  \centering
  \includegraphics[width=1\columnwidth]{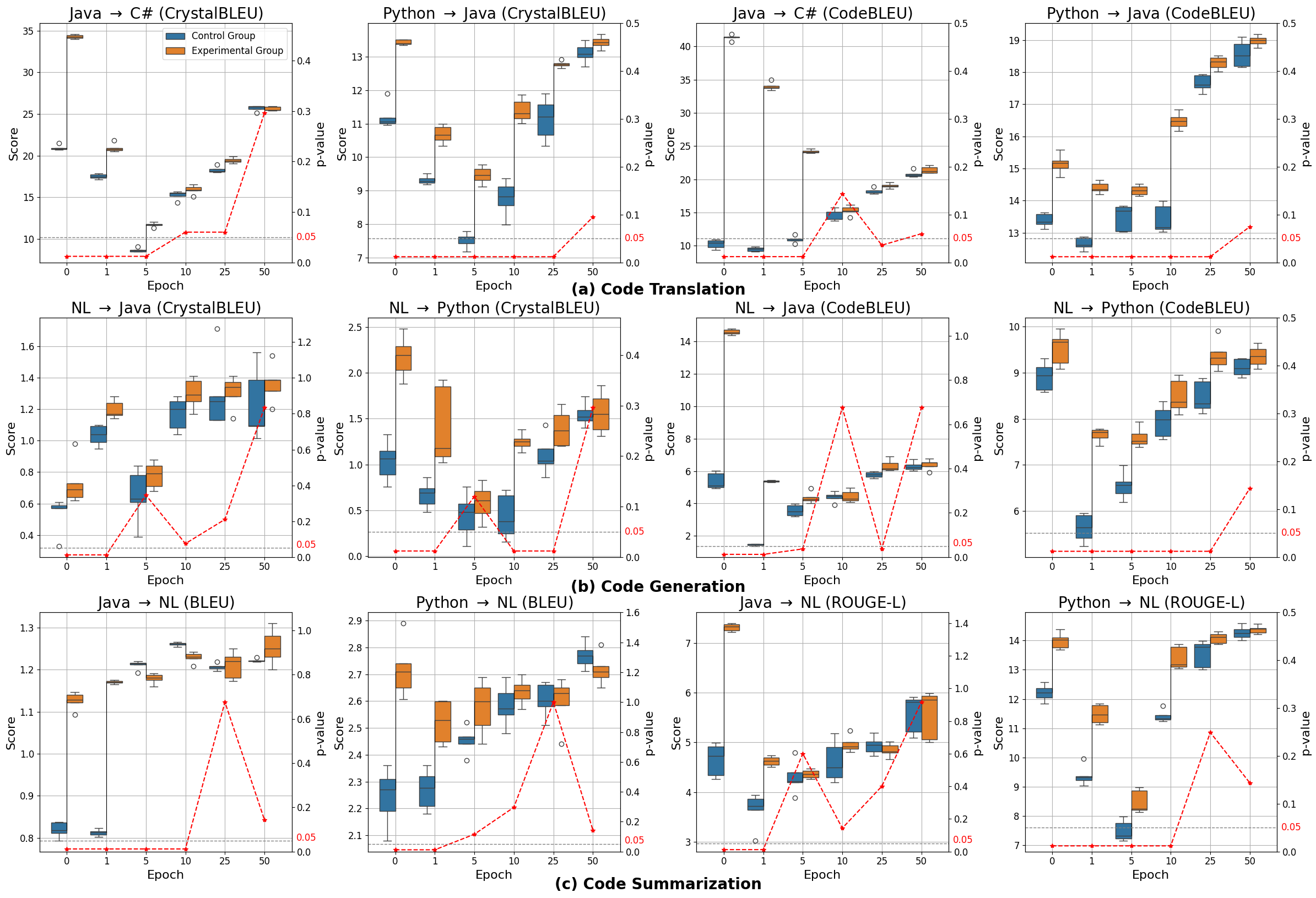}
  \caption{Performance Variation on Different Training Epochs}
  \label{fig:discuss}
\end{figure*}

To further investigate the influence of fine-tuning on contamination effects, towards each studied code task, we reduce the fine-tuning size to 500 samples and progressively enlarge the training epochs (\textit{E}=$\{$0, 1, 5, 10, 25, 50$\}$) for multiple experiments, thereby simulating the expansion of the fine-tuning scale and controlling the extent of learning underlying knowledge and task alignment specifications. As with our previous RQs, we repeat the experiment 5 times for each setting, and the results are presented in Fig.~\ref{fig:discuss}, where \textit{E} = 0 denotes zero-shot inference without fine-tuning. For each pair of control and experimental groups, we use a black line to connect them, thereby making their performance differences clearer. Meanwhile, \textit{p}-values for each experiment are plotted with red dots connected by dashed lines in Figure \ref{fig:discuss}. As can be seen, the contaminated experimental group overall performs significantly better than the uncontaminated counterpart initially (e.g., \textit{E}$\leq$5 for almost all settings).
With the training epochs enlarging, which means more underlying knowledge and task alignment specifications are learned by GPT-2, the contamination effects gradually eliminate. Besides, we also find an interesting phenomenon: the average performance of GPT-2 under both experimental and control groups shows a trend of declining first, then rising across almost all code tasks. A potential explanation is that the pre-training stage of our experiments mainly learn to generate Java/Python code given Java/Python prefixes, which is relatively different from the fine-tuning tasks, i.e., code translation, generation, and summarization. Thus, there are certain conflicts between them, leading to the re-learning of the task alignment on all downstream code tasks at the initial fine-tuning stage. With the training continuing, GPT-2 gradually overcomes the task alignment conflicts, thereby recovering and even improving its performance, which is also a phenomenon revealed in \cite{zhengspurious}.
\begin{boxK}
   \textbf{Finding 5:} 
   Large-scale fine-tuning eliminates the effect of paired contamination in decoder-only PLMs.
   In contrast, small-scale (e.g., 500 samples with fewer than 5 training epochs) or no fine-tuning uncovers this performance overestimation.
\end{boxK}

\begin{table}[htbp]
  \centering
  \setlength{\abovecaptionskip}{0cm}
  \setlength{\tabcolsep}{15pt}
    \caption{Results of Direct Inference Using Pre-trained GPT-2}
    \label{tab:6}
    \scriptsize
    \begin{tabular}{@{}c l c c@{}}
      \toprule
      \multirow{2}{*}{\textbf{Task}} 
      & \multirow{2}{*}{\textbf{In / Out}} 
      & \multicolumn{2}{c}{\textbf{Model Performance (Experimental Group / Control Group / \textit{p}-value)}} \\
      \cmidrule(lr){3-4}
      & & \textbf{Cry-BLEU} 
          & \textbf{CodeBLEU} \\
      \midrule

      \multirow{2}{*}{\makecell{Code Translation}}
      & Java$\rightarrow$C\#
      & \textbf{34.83}$_{\pm0.17}$ / 21.03$_{\pm0.16}$ / \textcolor{red}{0.012}
      & \textbf{41.48}$_{\pm0.80}$ / 10.00$_{\pm0.20}$ / \textcolor{red}{0.012} \\

      & Python$\rightarrow$Java
      & \textbf{13.21}$_{\pm0.20}$ / 11.43$_{\pm0.07}$ / \textcolor{red}{0.012}
      & \textbf{15.17}$_{\pm0.13}$ / 13.95$_{\pm0.07}$ / \textcolor{red}{0.012} \\

      \multirow{2}{*}{\makecell{Code Generation}}
      & NL$\rightarrow$Java
      & \textbf{0.72}$_{\pm0.06}$ / 0.63$_{\pm0.04}$ / \textcolor{red}{0.012}
      & \textbf{14.55}$_{\pm0.62}$ / 5.77$_{\pm0.19}$ / \textcolor{red}{0.012} \\

      & NL$\rightarrow$Python
      & \textbf{2.17}$_{\pm0.23}$ / 1.17$_{\pm0.22}$ / \textcolor{red}{0.012}
      & \textbf{9.45}$_{\pm0.23}$ / 9.10$_{\pm0.31}$ / \textcolor{red}{0.012} \\

        \midrule
        & & \textbf{BLEU} & \textbf{ROUGE-L} \\
        \midrule
      \multirow{2}{*}{\makecell{Code Summarization}}
      & Java$\rightarrow$NL
      & \textbf{1.13}$_{\pm0.02}$ / 0.83$_{\pm0.03}$ / \textcolor{red}{0.012}
      & \textbf{7.19}$_{\pm0.08}$ / 4.30$_{\pm0.04}$ / \textcolor{red}{0.012} \\

      & Python$\rightarrow$NL
      & \textbf{2.64}$_{\pm0.17}$ / 2.24$_{\pm0.11}$ / \textcolor{red}{0.012}
      & \textbf{13.99}$_{\pm0.28}$ / 12.21$_{\pm0.28}$ / \textcolor{red}{0.012} \\
      \bottomrule
    \end{tabular}
\end{table}

\section{Implications}


\textbf{Implications for Researchers:} 
As the first systematic study of fine-grained contamination effects across diverse PLMs/LLMs, PLs, and code tasks, our experimental/control group construction in Section \ref{Research Methodology} scales to massive LLM corpora (>1 trillion tokens) and is reproducible. 
Secondly, since most LLM code corpora come from Code Hosting Platforms (e.g., GitHub), for tasks whose samples are naturally unpaired or only partially occur (e.g., test generation, code translation), researchers need not keep rebuilding benchmarks around training cut-off dates—our evidence supports this and saves substantial effort.
Thirdly, for data contamination mitigation studies, our work reveals that researchers should focus on the fine-tuning phase if the PLM was fine-tuned before inference. However, for the pretrain–inference paradigm, the usage scenario of most LLMs, the pretraining stage should be concentrated.


\textbf{Implications for Practitioners:} Firstly, in Section \ref{Discussion}, we find that although PLMs suffer from contamination during pre-training, their performance does not show significant improvement on the test set after fine-tuning, indicating that the impact of contamination can be effectively mitigated to some extent. Therefore, deploying PLMs with fine-tuning in practice is safe for data contamination of the pre-training stage, contrary to the conclusion from NLP research \cite{jiang2024does} mentioned in Section \ref{RQ1: Does Input-Only Contamination Lead to Performance Overestimation?}. In other words, contamination in fine-tuning should be given more attention instead of pre-training for PLMs.
Secondly, when LLM practitioners are doing data cleaning before LLM pre-training \cite{touvron2023llama, jiang2025aixcoder}, they need not allocate efforts to code samples that are unpaired or only partially occur in follow-up benchmarks, because these kinds of contamination carry insignificant effects on downstream performance, as we revealed in Section \ref{Experimental Result}.


\section{Threats to Validity}
\textbf{Limited contamination scale:} In this study, the proportion of contaminated data in experiments of both PLMs and LLMs is relatively small compared to their pre-training corpora, without further exploration of the larger scale of contamination.
However, it is worth noting that in real-world scenarios, benchmark evaluation datasets, such as HumanEval \cite{chen2021evaluating} and ClassEval-T \cite{xue2025classeval}, contain only 164 Python functions and 282 class-level code pairs, aligning with our experimental settings on data volumes. Therefore, our experimental results remain reasonably credible for code intelligence practice despite this potential limitation.


\textbf{Perturbation quality:} When constructing uncontaminated datasets for LLMs, we manually perturb the contaminated data to be unseen by LLMs, leading to two threats concerning data quality: 1) whether the perturbed sample can be considered uncontaminated, and 2) whether the perturbation affects the sample complexities. Following most previous workarounds for avoiding data contamination \cite{rao2025codemorph, wang2023recode, sallou2024breaking}, we adopt a relatively high perturbation rate of 20.38\%--24.65\% on average across various experiments to craft LLM-unseen samples, shown in Tables~\ref{tab:data-rq1}--\ref{tab:data-rq4}.
To further eliminate the first threat, we performed an item-by-item comparison between the perturbed data and the publicly available pre-training corpora, finding no duplicate entries. 
As for the second threat, two authors of this paper participate in the perturbation with a double-checking mechanism to ensure the consistency of token lengths and cyclomatic complexity before and after perturbation. According to the dataset statistics listed in Tables~\ref{tab:data-rq1}--\ref{tab:data-rq4}, the sample complexities remain unchanged to a great extent. Thus, we believe the second threat is also minimized

\section{Related Work}
\subsection{Code Intelligence Tasks on PLMs and LLMs}
Recent years have witnessed extensive research on code intelligence tasks based on PLMs/LLMs, such as code generation \cite{li2022competition, le2022coderl}, code summarization \cite{ahmed2022few,geng2024large}, and code translation \cite{yang2024exploring,xue2025classeval}. Early efforts, such as CodeBERT~\cite{feng-etal-2020-codebert} and GraphCodeBERT \cite{guo2020graphcodebert}, jointly modeled NL and PL, enhancing semantic understanding of code through contrastive learning and structural awareness. Subsequently, PLBART~\cite{ahmad2021unified}, based on an encoder-decoder architecture, demonstrated strong transfer capabilities. CodeT5~\cite{wang2021codet5} was further pre-trained on T5 \cite{raffel2020exploring}, achieving state-of-the-art performance on a variety of code understanding and generation tasks. On the LLM front, Codex \cite{chen2021evaluating} and CodeGen \cite{nijkamp2022codegen} have shown remarkable capabilities in multi-turn reasoning and complex programming scenarios. More recent works, such as StarCoder \cite{li2023starcoder} and WizardCoder \cite{luo2023wizardcoder} explore the upper performance bounds of open-source LLMs in code intelligence, emphasizing the importance of model scale, instruction tuning, and multilingual training for improving code comprehension and generation abilities.

\subsection{Studies on Data Contamination}
Data contamination studies have been investigated for years; many are dedicated to contamination detection.
For instance, Golchin et al.~\cite{golchin2024time} proposed both instance-level and partition-level contamination detection, while Dong et al.~\cite{dong2024generalization} proposed CDD, using output distributions to detect contamination in LLMs on both code and NLP tasks. Deng et al. \cite{deng2024investigating}, instead, proposed TS-Guessing to detect contamination by examining whether LLMs can produce masked wrong answers.
As for research exploring contamination effects, early research was conducted in the NLP field. For example, Magar et al. \cite{magar2022data} explored to what extent BERT exploits or just memorises the contaminated data for downstream tasks. Jiang et al. \cite{jiang2024does} explored the input-only/output-only contamination effects on GPT-2. Subsequently, contamination has also been studied in the code intelligence field, Riddell et al. \cite{riddell2024quantifying}, and Ramos et al. \cite{ramos2024large} quantified the contamination effects on LLM-based code generation and bug fixing, respectively. Cao et al. \cite{cao2024concerneddatacontaminationassessing} extracted data before and after LLMs' cut-off date to study the effect of countermasures on coding tasks. Yang et al. \cite{yang2024unveiling} prompt code LLMs to generate massive code outputs and compare them with the corresponding pre-training corpora, thereby quantifying the memorization phenomenon for establishing taxonomy and detecting/mitigating related risks. Jain et al. \cite{jain2024livecodebench} propose LiveCodeBench, a continuously updated coding benchmark that mitigates contamination during evaluation. Al-Kaswan et al. \cite{alkaswan2024traces} leverage data extraction attacks to quantify memorization in code LLMs through code completion. Nie et al. \cite{nie2025desec} propose DESEC, a token-level characterization method for effectively extracting memorized secrets from code LLMs.

Nonetheless, owing to the distinctive features of code tasks, as we mentioned in Section \ref{RQ1: Does Input-Only Contamination Lead to Performance Overestimation?}, different conclusions from the NLP research concerning input/output-only contamination are obtained. 
Focusing on the domain of code intelligence, most studies only consider sample-level contamination \cite{cao2024concerneddatacontaminationassessing, riddell2024quantifying, ramos2024large}, i.e., the fourth setting in this paper, for detecting or evaluating memorized content. As such, the effects of other fine-grained contamination issues remain unanswered questions, which is a striking gap this paper fills. 

\section{Conclusion and Future Work}
This paper presents a fine-grained empirical study to reveal the impact of different data contamination scenarios on code intelligence. The whole experiment covers diverse PLMs/LLMs, PLs, and code tasks, uncovering different conclusions that were released before in the NLP field, and challenging the conventional belief that contamination inevitably induces performance overestimation in the code intelligence field, manifesting the necessity of our research and offering practical guidance for practitioners.
Future work could further explore the effects of larger-scale contamination and their influence on PLM/LLMs with post-training. 

\noindent\textbf{Data Availability:} We open-sourced the replication package and datasets of experimental and control groups in each contamination setting at \cite{anonymous:online}.

\begin{acks}
This work was partially supported by the National Natural Science Foundation of China (Grant Nos. 62502283, U24B20149, 62402482), the Natural Science Foundation of Shandong Province (Grant No. ZR2024QF093), and the Young Talent of Lifting Engineering for Science and Technology in Shandong, China (Grant No. SDAST2025QTB031).
\end{acks}


\bibliographystyle{ACM-Reference-Format}
\bibliography{sample-base}

\end{document}